\documentclass[aps,prd,preprint,groupedaddress,showpacs,amsmath]{revtex4}
\usepackage{graphicx}
\usepackage{bm}
\begin{document}


\title{Resonances Driven by a Neutrino Gyroscope and Collective
Neutrino Oscillations in Supernovae}


\author{Meng-Ru Wu}
\email{mwu@physics.umn.edu}
\author{Yong-Zhong Qian}
\email{qian@physics.umn.edu}
\affiliation{School of Physics and Astronomy, University of Minnesota, Minneapolis, MN 55455}

\date{\today}

\begin{abstract}
We show that flavor evolution of a system of neutrinos with continuous
energy spectra as in supernovae
can be understood in terms of the response of individual 
neutrino flavor-isospins (NFIS's) to the mean field. In the case of
a system initially consisting of $\nu_e$ and $\bar\nu_e$ with the
same energy spectrum but different number densities, the mean field
is very well approximated by the total angular momentum of a
neutrino gyroscope. Assuming that NFIS evolution is independent of
the initial neutrino emission angle, the so-called single-angle 
approximation, we find that the evolution is governed by two types of 
resonances driven by precession and nutation of
the gyroscope, respectively. The net flavor transformation crucially
depends on the adiabaticity of evolution through these resonances.
We show that the results for the system of two initial neutrino
species can be extended to a system of four species with the initial
number densities of $\nu_e$ and $\bar\nu_e$ significantly larger than those of
$\nu_x$ and $\bar\nu_x$. Further, we find that when the dependence on
the initial neutrino emission angle is taken into account in the multi-angle 
approximation, nutation of the mean field is quickly damped out and can be 
neglected. In contrast, precession-driven resonances still govern the evolution of
NFIS's with different energy and emission angles just as in the single-angle
approximation. Our pedagogical and analytic study of collective neutrino 
oscillations in supernovae provides some insights into these seemingly 
complicated yet fascinating phenomena.
\end{abstract}


\maketitle


\section{Introduction\label{sec-intro}}
Experiments on solar, atmospheric, reactor, and accelerator neutrinos
have found overwhelming evidence for neutrino flavor transformation 
via vacuum oscillations and the Mikheyev-Smirnov-Wolfenstein (MSW) 
effect (see \cite{PDG10} and references therein). The parameters 
governing these phenomena have largely been determined except for
the neutrino mass hierarchy, 
the vacuum mixing angle $\theta_{13}$, and the CP-violation phase. 
For supernova neutrinos, in addition to the MSW effect induced by 
neutrino forward scattering on electrons in matter \cite{Wolfenstein, MS}, 
novel phenomena of flavor transformation can occur 
(see \cite{ARNPS} and references therein)
due to neutrino forward scattering on other neutrinos
\cite{FMWS, Pantaleone, SR}. This is because near the proto-neutron star 
produced in a supernova, the number density of neutrinos can exceed
that of electrons. The dominance of neutrino-neutrino over neutrino-electron
forward scattering couples flavor evolution of neutrinos that have
different energy and travel on different trajectories,
thereby giving rise to new phenomena. Assuming no CP-violation in
the neutrino sector, \cite{0411159,0505240,0511275} 
first showed that collective oscillations of supernova neutrinos can
occur for the vacuum-mass-squared difference $\delta m_{13}^2$
relevant for atmospheric neutrino oscillations. This collective
flavor transformation can be 
treated effectively as $2\times 2$ mixing between $\nu_e$
and $\nu_x$ ($\bar\nu_e$ and $\bar\nu_x$), 
where $\nu_x$ ($\bar\nu_x$) is a superposition of
$\nu_\mu$ and $\nu_\tau$ ($\bar\nu_\mu$ and 
$\bar\nu_\tau$), because $\nu_\mu$ and $\nu_\tau$ ($\bar\nu_\mu$ and 
$\bar\nu_\tau$) have identical emission characteristics and interactions
in supernovae. 

Subsequent to the study in \cite{0511275}, 
a number of authors have investigated collective oscillations of supernova 
neutrinos (see \cite{ARNPS} for a review). The most
prominent feature of collective oscillations is the
stepwise swap of the $\nu_e$ and $\nu_x$ energy spectra
at some characteristic ``split'' energy \cite{0606616,0705.1830}.
As this spectral swap for supernova neutrinos depends on the neutrino mass 
hierarchy (i.e., the sign of $\delta m_{13}^2$) and $\theta_{13}$, 
neutrino signals from a Galactic supernova can potentially provide 
a sensitive probe of both these unknown parameters.

Recently, it was found that spectral swaps for supernova neutrinos can occur at more 
than one energy \cite{0707.1998,0904.3542,1012.1339}. Specifically, using
supernova neutrino emission parameters with initial number densities of 
$\nu_e$ and $\bar\nu_e$ significantly larger than those of $\nu_x$
and $\bar\nu_x$, \cite{0707.1998} found that for an inverted
mass hierarchy (IH, the lightest mass eigenstate being dominantly 
$\nu_x$), an additional swap occurs in the spectra of $\bar\nu_e$ 
and $\bar\nu_x$ at a rather low energy. In contrast, using very different parameters
with initial number densities of $\nu_x$ and $\bar\nu_x$ significantly larger 
than those of $\nu_e$ and $\bar\nu_e$, \cite{0904.3542} and \cite{1012.1339}
showed that spectral swaps occur at more than one energy for both IH and 
normal mass hierarchy (NH, the heaviest mass eigenstate being dominantly 
$\nu_x$). 

In this paper we focus on the occurrence of spectral swaps for the case where
the initial number densities of $\nu_e$ and $\bar\nu_e$ are significantly larger than
those of $\nu_x$ and $\bar\nu_x$. As a concrete example,
we assume that neutrinos are emitted in one of the flavor eigenstates
$\nu_a$ (representing any one of $\nu_e$, $\nu_x$, $\bar\nu_e$, and
$\bar\nu_x$) from a sharp neutrino sphere of radius $R_\nu=10$~km 
with a normalized spectrum of the form
\begin{equation}\label{eq-normspec}
f_{\nu_a}(E)=\frac{128}{3}\frac{E^3}{\langle E_{\nu_a}\rangle^4}
\exp(-4E/\langle E_{\nu_a}\rangle),
\end{equation} 
where $\langle E_{\nu_a}\rangle$ is the average energy. We take
$\langle E_{\nu_e}\rangle=10$~MeV,
$\langle E_{\bar\nu_e}\rangle=15$~MeV, and
$\langle E_{\nu_x}\rangle=\langle E_{\bar\nu_x}\rangle=24$~MeV.
We further take the neutrino luminosity to be
$L_{\nu_a}=10^{51}$~erg~s$^{-1}$ for each species so that the 
initial number densities of $\nu_e$, $\bar\nu_e$, $\nu_x$, and $\bar\nu_x$
are in ratios of $2.4:1.6:1:1$. 

We define $\delta m^2\equiv|\delta m^2_{13}|>0$ and $\theta_{\rm v}$ 
to be the parameters for vacuum mixing between $\nu_e$ and $\nu_x$ 
($\bar\nu_e$ and $\bar\nu_x$), where $0<\theta_{\rm v}<\pi/4$ 
or $\pi/4<\theta_{\rm v}<\pi/2$ corresponds to the NH or IH, respectively. 
In accordance with the experimental limits on $\theta_{13}$,
we take $\theta_{\rm v}\ll 1$ (NH) or 
$\tilde{\theta}_{\rm v}\equiv\pi/2-\theta_{\rm v}\ll 1$ (IH). 
Following the definition of \cite{0511275}, we represent each neutrino 
by a neutrino flavor isospin (NFIS) $\mathbf{s}$ of 
magnitude $1/2$ in a three-dimensional flavor space. In particular,
\begin{equation}\label{eq-nfis1}
\mathbf{s}=\begin{cases}
\mathbf{\hat{e}}_z^{\rm f}/2,&\mbox{for $\nu_e$ or $\bar\nu_x$},\\
-\mathbf{\hat{e}}_z^{\rm f}/2,&\mbox{for $\nu_x$ or $\bar\nu_e$},
\end{cases}
\end{equation}
where $\mathbf{\hat{e}}_z^{\rm f}$ is the unit vector
in the $z$-direction of the flavor space. It is convenient to use
\begin{equation}\label{eq-omega}
\omega\equiv\begin{cases}
\delta m^2/2E,&\mbox{for $\nu_e$ and $\nu_x$},\\
-\delta m^2/2E,&\mbox{for $\bar\nu_e$ and $\bar\nu_x$},
\end{cases}
\end{equation}
to label the NFIS of an initial $\nu_a$ with energy $E$ as 
$\mathbf{s}_\omega$. 

We follow the flavor evolution of neutrinos in the above example
from the neutrino sphere to a radius $r=250$~km, where oscillations have 
effectively ceased. We focus on the IH with $\delta m^2=3\times 10^{-3}$~eV$^2$
and assume $\tilde\theta_{\rm v}=10^{-5}$ in view of matter suppression of mixing,
but otherwise ignore the matter above the neutrino sphere.
We further assume that the flavor evolution of a $\nu_a$ emitted in 
a non-radial direction is identical to that of another $\nu_a$ with 
the same energy but emitted radially. This is the so-called 
``single-angle'' approximation.
The effective total number density at radius $r>R_\nu$ for neutrinos
emitted initially as $\nu_a$ is taken to be
\begin{equation}\label{eq-nr}
n_{\nu_a}(r)=\frac{L_{\nu_a}}
{4\pi R_\nu^2\langle E_{\nu_a}\rangle}
\left[1-\sqrt{1-(R_\nu/r)^2}\right]^2.
\end{equation}
The evolution of the NFIS $\mathbf{s}_\omega$ with time $t$, or equivalently,
with radius $r$ is governed by
\begin{equation}\label{eq-nfiseom}
\frac{d}{dt}\mathbf{s}_\omega=\frac{d}{dr}\mathbf{s}_\omega=
\mathbf{s}_\omega\times\left[
\omega\mathbf{H}_{\rm v}-2\sqrt{2}G_F\sum_an_{\nu_a}(r)\int_0^{\infty}
\mathbf{s}_{\omega'}f_{\nu_a}(E)dE\right],
\end{equation}
where $\mathbf{H}_{\rm v}\equiv -\sin2\theta_{\rm v}\mathbf{\hat{e}}_x^{\rm f} 
+\cos2\theta_{\rm v}\mathbf{\hat{e}}_z^{\rm f}=
-\sin2\tilde\theta_{\rm v}\mathbf{\hat{e}}_x^{\rm f} 
-\cos2\tilde\theta_{\rm v}\mathbf{\hat{e}}_z^{\rm f}$,
$G_F$ is the Fermi constant, and $\omega'$ runs over all NFIS's 
as summation over $a$ and integration over $E$ are performed.
For convenience, we adopt the natural units, where the Planck constant
$\hbar$ and the speed of light $c$ are set to unity, in writing expressions
and equations throughout the paper.

\begin{figure}
\includegraphics*[width=10cm, angle=270]{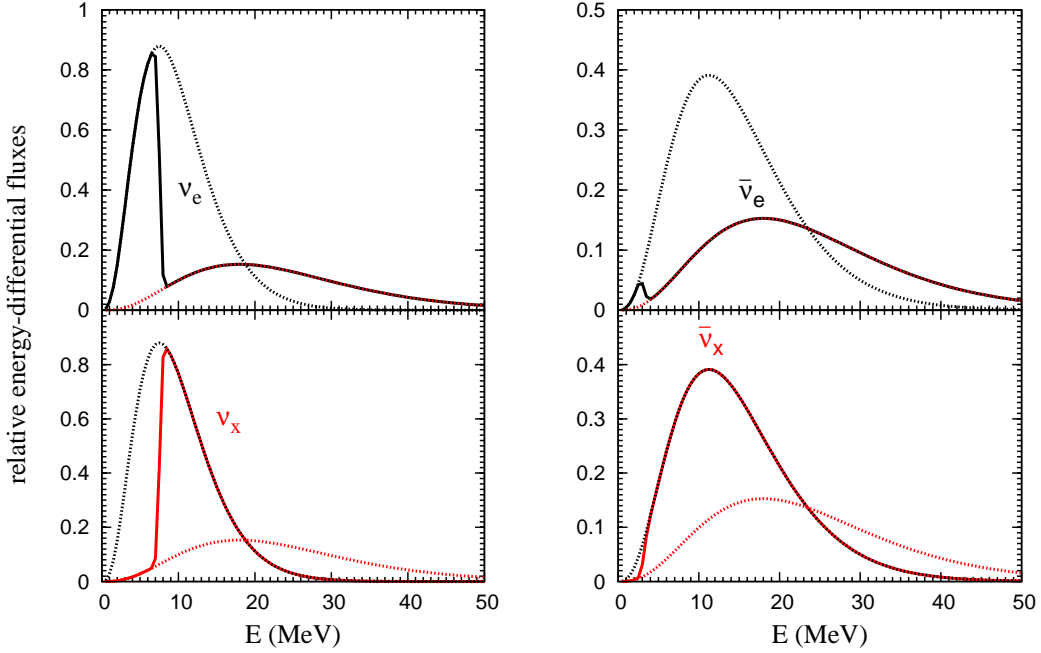}
\caption{An example of stepwise spectral swaps for the IH and for the case where the
initial number densities of $\nu_e$ and $\bar\nu_e$ are significantly larger
than those of $\nu_x$ and $\bar\nu_x$. The neutrino flux spectra at the neutrino sphere 
(dashed curves) and at $r=250$~km (solid curves) are shown. The dashed curves with
larger maxima are for the initial $\nu_e$ and $\bar\nu_e$ while those with lower maxima
are for the initial $\nu_x$ and $\bar\nu_x$.
It can be seen that the $\nu_e$ and $\nu_x$ spectra are swapped for
$E\gtrsim 7.6$~MeV and the $\bar\nu_e$ and $\bar\nu_x$ spectra are swapped
for $E\gtrsim 3.1$~MeV.\label{fig-splits}}
\end{figure}

\begin{figure}
\includegraphics*[width=60mm, angle=270]{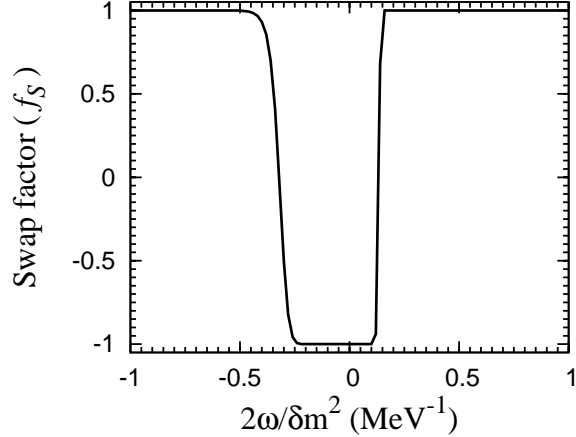}
\caption{The swap factor $f_S(\omega,t)$ at $r=250$~km as a function of 
$2\omega/\delta m^2$ for the example shown in Fig.~\ref{fig-splits}. Note
that the negative range of the horizontal axis corresponds to $\bar\nu_e$ and
$\bar\nu_x$ while the positive range corresponds to $\nu_e$ and $\nu_x$.
\label{fig-swaps}}
\end{figure}

Figure~\ref{fig-splits} compares the neutrino flux spectra at the neutrino sphere 
(dashed curves) and at $r=250$~km (solid curves). The vertical scale measures
the relative energy-differential fluxes, which are 
$\propto L_{\nu_a}f_{\nu_a}(E)/\langle E_{\nu_a}\rangle$ at the neutrino sphere.
It can be seen that the $\nu_e$ and $\nu_x$ spectra are swapped for
$E\gtrsim 7.6$~MeV and the $\bar\nu_e$ and $\bar\nu_x$ spectra are swapped
for $E\gtrsim 3.1$~MeV. To highlight these spectral changes, 
we define a ``swap factor''
\begin{equation}
f_S(\omega,t)\equiv s^{\rm f}_{\omega,z}(t)/s^{\rm f}_{\omega,z}(0),
\end{equation}
where $s^{\rm f}_{\omega,z}(t)\equiv \mathbf{s}_\omega(t)\cdot
{\mathbf{\hat e}}^{\rm f}_z$. The value of $f_S(\omega,t)=1$ or $-1$ corresponds 
to complete survival or conversion of the initial $\nu_a$, respectively.
The swap factor $f_S(\omega,t)$ at $r=250$~km for the above example
is shown as a function of $2\omega/\delta m^2$ in Fig.~\ref{fig-swaps}.

The main goal of this paper is to explain how spectral swaps occur in
the above example. Our approach is pedagogical. In Sec.~\ref{sec-meanfield} 
we consider the net effect of all the NFIS's in the above example as a ``mean field'' 
to which an individual $\mathbf{s}_\omega$ responds. We show that the evolution 
of this mean field can be well described by the motion of a ``neutrino gyroscope'' 
over a wide range of neutrino densities.
In Sec.~\ref{sec-gyro} we discuss in detail the precession and nutation of the
neutrino gyroscope. In Sec.~\ref{sec-res} we show that the response of an
NFIS to the mean field is characterized by two types of resonances driven 
by the precession and nutation of the neutrino gyroscope, respectively. These
resonances and the adiabaticity of evolution through them give rise
to the stepwise spectral swaps. In Sec.~\ref{sec-snnu}, we relax the 
single-angle approximation and discuss the evolution of NFIS's along different 
trajectories in realistic supernova environments. 
We show that the trajectory-dependent evolution strongly suppresses
nutation of the neutrino gyroscope. Consequently, the spectral swaps in this case 
are determined by precession-driven resonances only. We summarize our
results and give conclusions in Sec.~\ref{sec-conc}.

\section{Mean Field of NFIS's and Neutrino Gyroscope}
\label{sec-meanfield}
The NFIS for an initial $\nu_e$ and that for an initial $\nu_x$ with the same energy 
(or $\omega$) are equal in magnitude but opposite in direction [see Eq.~(\ref{eq-nfis1})].
It can be seen from Eq.~(\ref{eq-nfiseom}) that these two NFIS's remain equal in
magnitude but opposite in direction during their subsequent evolution. The same
is also true of the NFIS's for an initial $\bar\nu_e$ and an initial $\bar\nu_x$ with
the same energy. Hereafter, $\mathbf{s}_\omega$ for $\omega>0$ or $\omega<0$ 
refers to the NFIS for an initial $\nu_e$ or $\bar\nu_e$,
respectively. Then Eq.~(\ref{eq-nfiseom}) can be rewritten as
\begin{equation}\label{eq-nfis3}
\frac{d}{dr}\mathbf{s}_\omega=
\mathbf{s}_\omega\times\left[
\omega\mathbf{H}_{\rm v}-\mu(r)\int_{-\infty}^{\infty}
g(\omega')\mathbf{s}_{\omega'} d\omega'\right],
\end{equation}
where $\mu(r)\equiv 2\sqrt{2}G_Fn_{\nu_e}(r)$. In the above equation,
\begin{equation}\label{eq-gw}
g(\omega)=\frac{\delta m^2}{2\omega^2}\times
\begin{cases}
c_ef_{\nu_e}(E_\omega)-c_x f_{\nu_x}(E_\omega),&
\mbox{for $\omega>0$},\\
c_{\bar e}f_{\bar\nu_e}(E_\omega)-c_{\bar x}
f_{\bar\nu_x}(E_\omega),&\mbox{for $\omega<0$},
\end{cases}
\end{equation}
where $c_a=n_{\nu_a}(R_\nu)/n_{\nu_e}(R_\nu)=
\langle E_{\nu_a}\rangle/\langle E_{\nu_e}\rangle$ and 
$E_\omega=\delta m^2/2|\omega|$.

\begin{figure}
\includegraphics*[width=60mm, angle=270]{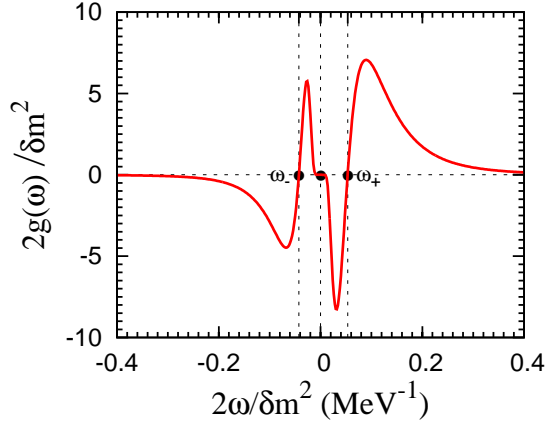}
\caption{Example supernova neutrino spectrum shown in terms of
$2g(\omega)/\delta m^2$ as a function of $2\omega/\delta m^2$.
There are three spectral crossings (filled circles)
at $\omega=\omega_-$, 0, and
$\omega_+$, respectively, for which $g(\omega)=0$.\label{fig-gw}}
\end{figure}

We define
\begin{equation}\label{eq-mfdef}
\mathbf{S}\equiv\int_{-\infty}^{\infty}g(\omega)\mathbf{s}_{\omega}d\omega
\end{equation}
and consider it as the mean field representing the net effect of all NFIS's
to which an individual $\mathbf{s}_{\omega}$ responds. Clearly, an exact
description of the mean field requires solving Eq.~(\ref{eq-nfis3}) for
all NFIS's. However, we can give an approximate description based on
the evolution of a small number of NFIS's. To see this, we show
$2g(\omega)/\delta m^2$ as a function of $2\omega/\delta m^2$
for our supernova example in Fig.~\ref{fig-gw}. There are three
``spectral crossings'' \cite{0904.3542} at $\omega=\omega_-$, 0, and
$\omega_+$, respectively, for which $g(\omega)=0$.
In each of the four spectral regions separated by these crossings,
the magnitude of $g(\omega)$ is large only for a relatively narrow range
of $\omega$.

We consider
that the four spectral regions $(-\infty,\omega_-)$, $(\omega_-,0)$,
$(0,\omega_+)$, and $(\omega_+,\infty)$ can be 
represented by four effective NFIS's $\mathbf{s}_{\bar e}^{(0)}$,
$\mathbf{s}_{\bar x}^{(0)}$, $\mathbf{s}_{x}^{(0)}$, and
$\mathbf{s}_{e}^{(0)}$, respectively, which correspond to a $\bar\nu_e$, 
a $\bar\nu_x$, a $\nu_x$, and a $\nu_e$ at the neutrino sphere.
We then approximate
$\mathbf{S}$ by
\begin{equation}
\mathbf{S}^{(0)}=
\alpha_{e}\mathbf{s}_{e}^{(0)}+
\alpha_{x}\mathbf{s}_{x}^{(0)}+
\alpha_{\bar x}\mathbf{s}_{\bar x}^{(0)}+
\alpha_{\bar e}\mathbf{s}_{\bar e}^{(0)},
\end{equation}
where the evolution of each $\mathbf{s}_{a}^{(0)}$ is governed by
\begin{equation}\label{eq-mfa2}
\frac{d}{dr}{\mathbf{s}}_{a}^{(0)}=\mathbf{s}_{a}^{(0)}\times
\left[\omega_{a}\mathbf{H}_{\rm v}-\mu(r)\mathbf{S}^{(0)}\right].
\end{equation}
In the above equations,
\begin{align}
\alpha_e&=\int_{\omega_+}^\infty |g(\omega)|d\omega,\\
\omega_e&=\frac{1}{\alpha_e}\int_{\omega_+}^\infty\omega |g(\omega)|d\omega,
\end{align}
and other quantities are defined similarly. For our supernova example,
$\alpha_{e}=0.78$, $\alpha_{x}=0.20$, $\alpha_{\bar x}=0.10$, and
$\alpha_{\bar e}=0.35$, while $2\omega_a/\delta m^2=0.15$, 0.032, $-0.027$,
and $-0.11$~MeV$^{-1}$ for $\mathbf{s}_{e}^{(0)}$, $\mathbf{s}_{x}^{(0)}$,
$\mathbf{s}_{\bar x}^{(0)}$, and $\mathbf{s}_{\bar e}^{(0)}$, respectively.

\begin{figure}
\includegraphics*[width=60mm, angle=270]{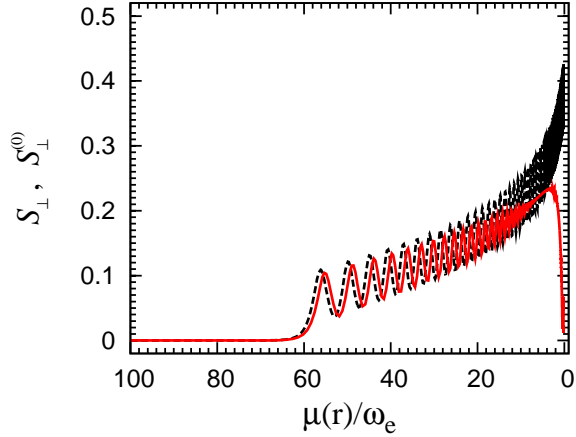}
\caption{Comparison of
$S_\perp$ (solid curve) and $S^{(0)}_\perp$ (dashed curve)
as functions of $\mu(r)/\omega_e$ for the supernova example.\label{fig-mfs}}
\end{figure}

At the neutrino sphere, $\mathbf{S}=\mathbf{S}^{(0)}={\mathbf{\hat e}}^{\rm f}_z/6$.
As can be shown from Eqs.~(\ref{eq-nfis3}) and (\ref{eq-mfa2}), the components 
of $\mathbf{S}$ and $\mathbf{S}^{(0)}$ parallel to $\mathbf{H}_{\rm v}$
are conserved during the subsequent evolution.
We numerically obtain their components perpendicular to $\mathbf{H}_{\rm v}$,
$S_\perp$ (solid curve) and $S^{(0)}_\perp$ (dashed curve),
and show these as functions of $\mu(r)/\omega_e$ in Fig.~\ref{fig-mfs}.
It can be seen that $S^{(0)}_\perp$ closely tracks $S_\perp$ for
$\mu(r)/\omega_e\gtrsim 20$ but large deviations occur for 
$\mu(r)/\omega_e<10$.
In particular, $S^{(0)}_\perp$ diverges from $S_\perp$ at $\mu(r)/\omega_e<4$.
The above results can be understood by comparing $\mu(r)|\mathbf{S}|$ with
the spread in $\omega$ for the spectrum shown in Fig.~\ref{fig-gw}. 
For $\mu(r)/\omega_e\gtrsim 20$, $\mu(r)|\mathbf{S}|\gtrsim 3\omega_e$ exceeds 
the spread over the entire spectrum, which is $\sim 2\omega_e$. So all NFIS's
evolve collectively in this regime and the four effective NFIS's included in
$\mathbf{S}^{(0)}$ are sufficient to give a good description of $\mathbf{S}$. 
For $\mu(r)/\omega_e\sim 10$, $\mu(r)|\mathbf{S}|\sim\omega_e$ approaches
$\Delta\omega_e=0.66\omega_e$, which is the spread in the spectral region
$(\omega_+,\infty)$ calculated from
\begin{equation}
\Delta\omega_e=\frac{1}{\alpha_e}\left[\int_{\omega_+}^\infty
(\omega-\omega_e)^2|g(\omega)|d\omega\right]^{1/2}.
\end{equation}
Consequently, for $\mu(r)/\omega_e<10$, 
the NFIS's in the above spectral region are no longer well
represented by $\mathbf{s}_{e}^{(0)}$ and large differences between
$S^{(0)}_\perp$ and $S_\perp$ occur. Eventually,
none of the effective NFIS's can represent their respective
spectral regions and $S^{(0)}_\perp$ diverges from $S_\perp$.

In a formal approach, we can use $\mathbf{S}^{(0)}$ as the zeroth order
approximation for $\mathbf{S}$ to solve Eq.~(\ref{eq-nfis3}) for the evolution of 
$\mathbf{s}_\omega$, and use the results to obtain a better approximation
for $\mathbf{S}$ from Eq.~(\ref{eq-mfdef}). This procedure may be repeated
until successive approximations for $\mathbf{S}$ converge. While this
approach does not save numerical efforts compared with solving Eq.~(\ref{eq-nfis3})
directly, it motivates an analytic study based on the zeroth order 
mean field $\mathbf{S}^{(0)}$, especially when $\mathbf{S}^{(0)}$ can be
understood with simple models. We carry out such a study in the rest of
the paper. We first consider a simpler case and then apply the results from this
case to discuss the supernova example in Sec.~\ref{sec-snnu}.

\subsection{System Initially Consisting of $\nu_e$ and $\bar\nu_e$ Only
\label{sec-nueanue}}
In our supernova example, the initial number densities of $\nu_e$ and $\bar\nu_e$
are significantly larger than those of $\nu_x$ and $\bar\nu_x$. To facilitate an
analytic study, we consider a simpler system initially consisting of $\nu_e$ and 
$\bar\nu_e$ only. We take normalized emission spectra of the form in 
Eq.~(\ref{eq-normspec}) 
with $\langle E_{\nu_e}\rangle=\langle E_{\bar\nu_e}\rangle=12$~MeV and
luminosities $L_{\nu_e}=1.2\times 10^{51}$ and 
$L_{\bar\nu_e}=0.8\times 10^{51}$~erg~s$^{-1}$ so that the initial number 
densities of $\nu_e$ and $\bar\nu_e$ are the same as
those in the supernova example. In this case, the effective neutrino spectrum
$g(\omega)$ reduces to
\begin{equation}\label{eq-hw}
h(\omega)=\frac{\delta m^2}{2\omega^2}\times
\begin{cases}
f_{\nu_e}(E_\omega),&
\mbox{for $\omega>0$},\\
\alpha f_{\bar\nu_e}(E_\omega),&\mbox{for $\omega<0$},
\end{cases}
\end{equation}
where $\alpha=n_{\bar\nu_e}(R_\nu)/n_{\nu_e}(R_\nu)=2/3$.

\begin{figure}
\includegraphics*[width=60mm, angle=-90]{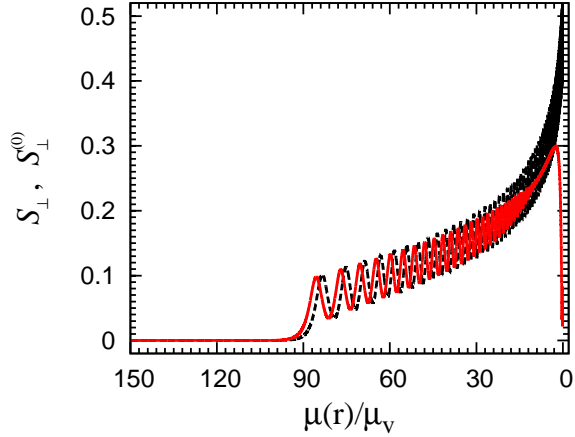}
\caption{Comparison of 
$S_\perp$ (solid curve) and $S^{(0)}_\perp$ (dashed curve)
as functions of $\mu(r)/\mu_{\rm v}$ for the system initially consisting of
$\nu_e$ and $\bar\nu_e$ only.\label{fig-sperp}}
\end{figure}

We consider that the zeroth order approximation for the mean field 
$\mathbf{S}=\int_{-\infty}^\infty h(\omega)\mathbf{s}_\omega d\omega$ 
is given by
\begin{equation}
\mathbf{S}^{(0)}\equiv\mathbf{s}_1^{(0)}+\alpha\mathbf{s}_2^{(0)},
\end{equation}
where $\mathbf{s}_1^{(0)}$ and $\mathbf{s}_2^{(0)}$ are the NFIS's
for an initial $\nu_e$ and an initial $\bar\nu_e$ with
\begin{subequations}
\begin{align}
\omega_1=\int_0^\infty\omega h(\omega) d\omega,\\
\omega_2=\frac{1}{\alpha}\int_{-\infty}^0\omega h(\omega) d\omega,
\end{align}
\end{subequations}
respectively. Specifically, $\omega_1=-\omega_2\equiv\mu_{\rm v}$ and
$2\mu_{\rm v}/\delta m^2=1/9$~MeV$^{-1}$. We show $S_\perp$ (solid curve)
and $S^{(0)}_\perp$ (dashed curve) as functions of $\mu(r)/\mu_{\rm v}$ in 
Fig. \ref{fig-sperp}. It can be seen that the comparison between $S_\perp$
and $S^{(0)}_\perp$ is very similar to that in the supernova example
[note that the horizontal scales for Figs.~\ref{fig-mfs} and \ref{fig-sperp}
are related by $\mu(r)/\mu_{\rm v}=1.35\mu(r)/\omega_e$]. 

\begin{figure}
\includegraphics*[width=60mm, angle=-90]{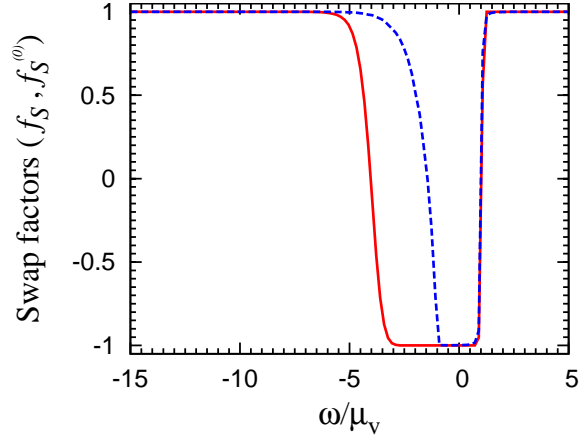}
\caption{Comparison of the swap factors
$f_S(\omega,t)$ (solid curve) and $f_S^{(0)}(\omega,t)$ (dashed curve)
at $r=250$~km for the system initially consisting of
$\nu_e$ and $\bar\nu_e$ only.\label{fig-swapcomp}}
\end{figure}

The swap factor $f_S(\omega,t)$ at $r=250$~km for the system initially 
consisting of $\nu_e$ and $\bar\nu_e$ only is shown as the solid curve
in Fig.~\ref{fig-swapcomp}. The swap factor $f_S^{(0)}(\omega,t)$
calculated from
\begin{equation}
\frac{d}{dr}{\mathbf{s}}_\omega=\mathbf{s}_\omega\times
\left[\omega\mathbf{H}_{\rm v}-\mu(r)\mathbf{S}^{(0)}\right],
\end{equation}
which uses $\mathbf{S}^{(0)}$ to approximate $\mathbf{S}$, is
shown as the dashed curve. It can be seen that just as in the supernova
example, $f_S(\omega,t)$ and $f_S^{(0)}(\omega,t)$ have two
characteristic split energies, one in the region of $\omega>0$ and the
other in the region of $\omega<0$. Although $f_S^{(0)}(\omega,t)$
has a different split energy from that of $f_S(\omega,t)$ in the region 
of $\omega<0$, they have the same qualitative behavior, especially the 
same split energy in the region of $\omega>0$. Our goal is to understand
the behavior of $f_S^{(0)}(\omega,t)$ analytically.

\subsection{Neutrino Gyroscope as the Approximate Mean Field\label{sec-gyromf}}
For the system initially consisting of $\nu_e$ and $\bar\nu_e$ only,
the evolution of $\mathbf{s}_1^{(0)}$ and $\mathbf{s}_2^{(0)}$ is 
governed by
\begin{subequations}
\begin{align}
\frac{d}{dr}{\mathbf{s}}_1^{(0)}&=\mathbf{s}_1^{(0)}\times
\left[\mu_{\rm v}\mathbf{H}_{\rm v}-\mu(r)\mathbf{S}^{(0)}\right],\\
\frac{d}{dr}{\mathbf{s}}_2^{(0)}&=\mathbf{s}_2^{(0)}\times
\left[-\mu_{\rm v}\mathbf{H}_{\rm v}-\mu(r)\mathbf{S}^{(0)}\right].
\end{align}
\end{subequations}
For convenience, we will drop the superscript ``(0)''
but otherwise use the symbols in the same meaning as in the above equations.
It is useful to consider the time evolution of $\mathbf{s}_1$ and $\mathbf{s}_2$ 
at a constant $\mu$ governed by
\begin{subequations}
\begin{align}
\mathbf{\dot s}_1&\equiv\frac{d}{dt}\mathbf{s}_1=
\mathbf{s}_1\times(\mu_{\rm v}\mathbf{H}_{\rm v}-\mu\mathbf{S}),\label{eq-s1}\\
\mathbf{\dot s}_2&\equiv\frac{d}{dt}\mathbf{s}_2=
\mathbf{s}_2\times(-\mu_{\rm v}\mathbf{H}_{\rm v}-\mu\mathbf{S}).\label{eq-s2}
\end{align}
\end{subequations}
As discussed in \cite{0608695,0703776} and repeated below, 
the system governed by the 
above equations is mathematically equivalent to a gyroscope in a uniform
gravitational field.

\begin{figure}
\includegraphics*[width=120mm, angle=0]{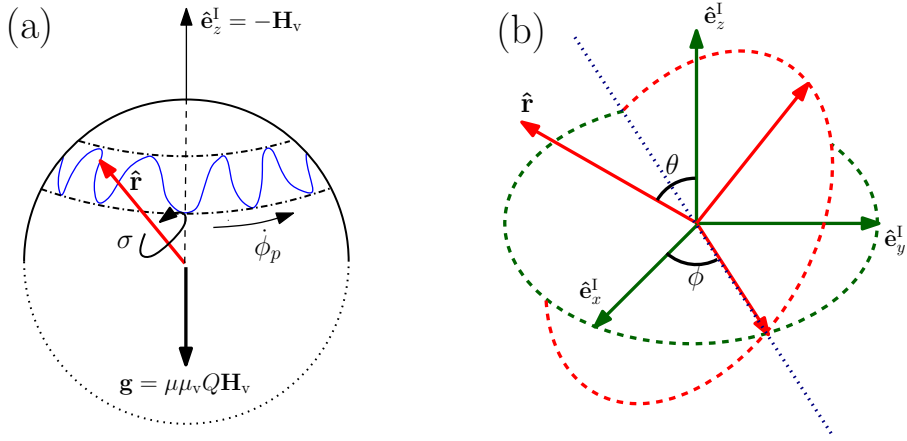}
\caption{Illustration of (a) the neutrino gyroscope and (b) the Euler angles
$\theta$ and $\phi$ used to describe its motion in Frame I. The gyroscope 
has a spinning point particle of mass $1/\mu$ attached to the
tip of a massless rod of unit length. The spin is along the direction 
$\mathbf{\hat r}$ of the rod and has a magnitude $\sigma$. The gyroscope 
is in a uniform gravitational field with an acceleration of gravity 
$\mathbf{g}$. The wiggly curve on the spherical surface crudely
indicates the trajectory of the tip of $\mathbf{\hat r}$ as the gyroscope
executes precession (change in $\phi$, arrow labeled by $\dot\phi_p$) 
and nutation (change in $\theta$) at a constant $\mu$.\label{fig-euler}}
\end{figure}

From Eqs.~(\ref{eq-s1}) and (\ref{eq-s2}) it is straightforward to show that
\begin{subequations}
\begin{align}
\mathbf{\dot Q}&=\mu\mathbf{S}\times\mathbf{Q},\label{eq-qg}\\
\mathbf{\dot S}&=\mu_{\rm v}\mathbf{Q}\times\mathbf{H}_{\rm v},
\label{eq-sg}
\end{align}
\end{subequations}
where
\begin{equation}\label{eq-defQ}
\mathbf{Q}=\mathbf{s}_1-\alpha\mathbf{s}_2+\frac{\mu_{\rm v}}{\mu}
\mathbf{H}_{\rm v}.
\end{equation}
From Eq.~(\ref{eq-qg}) it can be shown that $Q\equiv|\mathbf{Q}|$
is conserved. With the definition of a unit vector
$\mathbf{\hat r}\equiv\mathbf{Q}/Q$,
Eqs.~(\ref{eq-qg}) and (\ref{eq-sg}) can be rewritten as
\begin{subequations}
\begin{align}
\mathbf{S}&=\frac{1}{\mu}\mathbf{\hat r}\times\mathbf{\dot{\hat r}}+
\sigma\mathbf{\hat r},\label{eq-sgy}\\
\mathbf{\dot S}&=\frac{1}{\mu}\mathbf{\hat r}\times\mathbf{g},
\label{eq-stq}
\end{align}
\end{subequations}
where $\sigma\equiv\mathbf{S}\cdot\hat{\mathbf{r}}$ and
$\mathbf{g}\equiv\mu\mu_{\rm v}Q\mathbf{H}_{\rm v}$.
The above equations describe the motion of
a gyroscope in a uniform gravitational field with an
acceleration of gravity $\mathbf{g}$ (see Fig.~\ref{fig-euler}a). 
The gyroscope has a 
spinning point particle of mass $1/\mu$ attached to the
tip of a massless rod of unit length. 
The spin of the gyroscope is along the direction 
$\mathbf{\hat r}$ of the rod and it can be shown
from Eqs.~(\ref{eq-qg}) and (\ref{eq-sg}) that
the magnitude $\sigma$ of the spin is conserved. 
As can be seen from Eq.~(\ref{eq-stq}),
the component of the total angular momentum $\mathbf{S}$ 
parallel to $\mathbf{H}_{\rm v}$ is also conserved.

If $\mu$ varies smoothly with time, the neutrino gyroscope described above
evolves through a series of configurations corresponding to a continuous
range of $\mu$. The motion of this gyroscope provides 
a well-studied mechanical analog to the evolution of the approximate 
mean field.

\section{Precession and Nutation of The Neutrino Gyroscope
\label{sec-gyro}}
We specify the unit vector $\mathbf{\hat r}$ of the neutrino gyroscope by 
two of the Euler angles, $\theta$ and $\phi$, in Frame I: 
\begin{equation}
\mathbf{\hat r}=\sin\theta\sin\phi\,\mathbf{\hat e}_x^{\rm I}
-\sin\theta\cos\phi\,\mathbf{\hat e}_y^{\rm I}+
\cos\theta\,\mathbf{\hat e}_z^{\rm I},
\end{equation}
where $\mathbf{\hat e}_x^{\rm I}$, $\mathbf{\hat e}_y^{\rm I}$,
and $\mathbf{\hat e}_z^{\rm I}$ are the unit vectors associated
with the three axes of Frame I and 
$\mathbf{\hat e}_z^{\rm I}=-\mathbf{H}_{\rm v}$ (see Fig.~\ref{fig-euler}). 
We focus on the IH with $\tilde\theta_{\rm v}\ll 1$, for which 
$\mathbf{H}_{\rm v}=-\sin2\tilde\theta_{\rm v}\mathbf{\hat{e}}_x^{\rm f} 
-\cos2\tilde\theta_{\rm v}\mathbf{\hat{e}}_z^{\rm f}
\approx -{\mathbf{\hat e}}^{\rm f}_z$. Thus,
$\mathbf{\hat e}_z^{\rm I}\approx {\mathbf{\hat e}}^{\rm f}_z$ is the upward
direction and $\mathbf{g}$ (in the same direction as $\mathbf{H}_{\rm v}$)
points downward (see Fig.~\ref{fig-euler}a). 
When the neutrino gyroscope starts at $\mu\gg\mu_{\rm v}$, 
$\mathbf{Q}\approx\mathbf{s}_1-\alpha\mathbf{s}_2$ points in the direction of
${\mathbf{\hat e}}^{\rm f}_z$, i.e., it is in the upright position. This initial
configuration can give rise to interesting subsequent evolution.

\subsection{Motion of the Neutrino Gyroscope at Constant $\mu$\label{sec-cnnu}}
We first discuss motion of the neutrino gyroscope at constant $\mu$. 
Using Eq.~(\ref{eq-sgy}), we can express $\mathbf{S}$
in terms of $\theta$ and $\phi$ as
\begin{subequations}
\begin{align}
S_x&=\frac{1}{\mu}(\dot\theta\cos\phi-\dot\phi\sin\theta\cos\theta\sin\phi)
+\sigma\sin\theta\sin\phi,\label{eq-sx}\\
S_y&=\frac{1}{\mu}(\dot\theta\sin\phi+\dot\phi\sin\theta\cos\theta\cos\phi)
-\sigma\sin\theta\cos\phi,\label{eq-sy}\\
S_z&=\frac{1}{\mu}\dot\phi\sin^2\theta+\sigma\cos\theta
=\frac{1-\alpha}{2}\cos2\tilde\theta_{\rm v},\label{eq-sz}
\end{align}
\end{subequations}
where we have used conservation of $S_z$ in the exact form
(i.e., no approximation made for $\tilde\theta_{\rm v}\ll 1$) 
in the last equation. In addition to $\sigma$ and $S_z$, the third 
conserved quantity of the gyroscope is its total energy:
\begin{subequations}
\begin{align}
E_{\rm gyro}&=\frac{\mu}{2}\mathbf{S}^2-
\frac{1}{\mu}\mathbf{\hat r}\cdot\mathbf{g},\label{eq-egy}\\
&=\frac{1}{2\mu}\left(\dot\theta^2+\dot\phi^2\sin^2\theta\right)
+\frac{\mu}{2}\sigma^2+\mu_{\rm v}Q\cos\theta.\label{eq-etp}
\end{align}
\end{subequations} 
Conservation of $E_{\rm gyro}$ can be shown using 
Eqs.~(\ref{eq-sgy}) and (\ref{eq-stq}).
We can also derive an explicit equation of motion
from Eq.~(\ref{eq-stq}):
\begin{equation}\label{eq-eom}
\ddot\theta-\dot\phi(\dot\phi\cos\theta-\mu\sigma)\sin\theta=
\mu\mu_{\rm v}Q\sin\theta.
\end{equation}

In general, both $\theta$ and $\phi$ of a gyroscope evolve with time, 
and the corresponding motion (see Fig.~\ref{fig-euler})
is referred to as nutation ($\theta$)
and precession ($\phi$). For the neutrino gyroscope,
conservation of $\sigma$, $S_z$, $E_{\rm gyro}$ can be combined to 
give a conserved effective energy associated with nutation only:
\begin{equation}\label{eq-egy_eff}
E_\theta=E_{\rm gyro}-\frac{\mu}{2}\sigma^2
=\frac{\dot\theta^2}{2\mu}+V_{\rm eff}(\theta),
\end{equation}
where
\begin{equation}\label{eq-veff}
V_{\rm eff}(\theta)=\frac{\mu}{2}
\frac{(S_z-\sigma\cos\theta)^2}{\sin^2\theta}+
\mu_{\rm v}Q\cos\theta.
\end{equation}
As shown in Sec.~\ref{sec-vnnu}, nutation of the neutrino gyroscope
mostly occurs around the minimum of
the effective potential $V_{\rm eff}(\theta)$.
This potential minimum corresponds to $\theta=\theta_p$,
for which
\begin{subequations}
\begin{align}
\left.\frac{dV_{\rm eff}}{d\theta}\right|_{\theta_p}&=
\mu\frac{(S_z-\sigma\cos\theta_p)(\sigma-S_z\cos\theta_p)}
{\sin^3\theta_p}-\mu_{\rm v}Q\sin\theta_p=0,\label{eq-dvdt}\\
\mu\left.\frac{d^2V_{\rm eff}}{d\theta^2}\right|_{\theta_p}&=
\dot\phi_p^2\sin^2\theta_p+\left( 2\dot\phi_p\cos\theta_p - 
\mu\sigma\right)^2 \equiv\omega_n^2>0.\label{eq-d2vdt}
\end{align}
\end{subequations}
In Eq.~(\ref{eq-d2vdt}), $\dot\phi_p$ is the instantaneous precession 
frequency at the potential minimum and can be obtained from 
Eq.~(\ref{eq-sz}) after $\theta_p$ is solved from Eq.~(\ref{eq-dvdt}).
Note that Eq.~(\ref{eq-dvdt}) is the same as
\begin{equation}\label{eq-pre}
\dot\phi_p(\dot\phi_p\cos\theta_p-\mu\sigma)+\mu\mu_{\rm v}Q=0,
\end{equation}
which can also be obtained by setting $\ddot\theta=0$ and
$\dot\phi=\dot\phi_p$ in Eq.~(\ref{eq-eom}). In other words,
the minimum $V_{\rm eff}$ corresponds to
the maximum $\dot\theta$ [due to conservation of $E_\theta$,
see Eq.~(\ref{eq-egy_eff})], and hence $\ddot\theta=0$.

To the leading order, nutation can be approximated as oscillations of
$\theta$ around $\theta_p$ in response to the potential
\begin{equation}
V_{\rm eff}(\theta)\approx V_{\rm eff}(\theta_p)
+\frac{\omega_n^2}{2\mu}(\theta-\theta_p)^2.
\end{equation}
In this approximation, the evolution of $\theta$ can be described by
\begin{subequations}
\begin{align}
\theta&\approx\theta_p-\eta\cos(\omega_nt+\beta),\label{eq-theta}\\
\dot\theta&\approx\eta\omega_n\sin(\omega_nt+\beta),\label{eq-dottheta}
\end{align}
\end{subequations}
where $\eta$ is the amplitude of oscillation (or nutation)
and $\beta$ is a constant phase.
The values of $\eta$ and $\beta$ are determined by the initial
values of $\theta$ and $\dot\theta$ at $t=0$.
Using Eq.~(\ref{eq-sz}) to relate $\dot\phi$ at $\theta$ to 
$\dot\phi_p$ at $\theta_p$, we obtain to the first order in $\eta$,
\begin{subequations}
\begin{align}
\dot\phi&\approx\dot\phi_p-\eta\left[\frac{\mu\sigma-
2\dot\phi_p\cos\theta_p}{\sin\theta_p}\right]
\cos(\omega_nt+\beta),\label{eq-dotphi}\\
\phi&\approx\gamma+\dot\phi_pt-\eta\left[\frac{\mu\sigma-
2\dot\phi_p\cos\theta_p}{\omega_n\sin\theta_p}\right]
\sin(\omega_nt+\beta),\label{eq-phi}\end{align}
\end{subequations} 
where $\gamma$ is a constant determined by the 
initial value of $\phi$ at $t=0$.

From Eqs.~(\ref{eq-sx})--(\ref{eq-sz}) and 
(\ref{eq-theta})--(\ref{eq-phi}) we obtain 
\begin{align}
S_x+iS_y&\approx i\left\{\left(\frac{\dot\phi_p}{\mu}\cos\theta_p
-\sigma\right)\sin\theta_p+\frac{\eta}{2}\frac{\dot\phi_p}{\mu}
\left[\left(1-\frac{2\dot\phi_p-\mu S_z}{\omega_n}\right)
e^{i(\omega_nt+\beta)}\right.\right.\nonumber\\
&+\left.\left.\left(1+\frac{2\dot\phi_p-\mu S_z}{\omega_n}\right)
e^{-i(\omega_nt+\beta)}\right]\right\}e^{i(\dot\phi_pt+\gamma)},
\label{eq-sxy}
\end{align}
which is accurate to the first order in $\eta$.
The above equation shows that to the leading order, the angular
momentum $\mathbf{S}$ can be described by three precessing vectors
with different amplitudes and different precession frequencies,
which are $\dot\phi_p$, $\dot\phi_p+\omega_n$, and $\dot\phi_p-\omega_n$,
respectively. This leading-order expression of $\mathbf{S}$ provides
a very useful analytic description of the motion of the neutrino gyroscope.
Without loss of generality, hereafter we set the phase constants
$\beta=\gamma=0$ in Eq.~(\ref{eq-sxy}).

\subsection{Motion of the Neutrino Gyroscope for Slowly Decreasing $\mu$
\label{sec-vnnu}}
\begin{figure}
\includegraphics*[width=60mm, angle=-90]{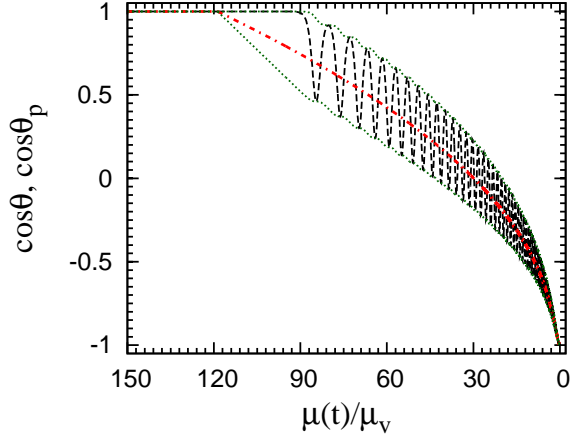}
\caption{Nutation of the neutrino gyroscope shown in terms of
$\cos\theta$ as a function of $\mu(t)/\mu_{\rm v}$ (dashed curve)
for $\tilde\theta_{\rm v}=10^{-5}$ (IH). The gyroscope is characterized
by $2\mu_{\rm v}/\delta m^2=1/9$~MeV$^{-1}$, $\alpha=2/3$, and
$\mu(t)=\mu(0)\left\{1-\sqrt{1-[\tau/(t+\tau)]^2}\right\}^2$, where
$\mu(0)/\mu_{\rm v}=2.52\times 10^5$ and $\tau=8.45/\mu_{\rm v}$.
The dot-dashed curve is for $\cos\theta_p$, which corresponds to the
minimum of the instantaneous $V_{\rm eff}(\theta)$, and the dotted 
curves are for $\cos\theta_{\rm max}$ and $\cos\theta_{\rm min}$,
which correspond to $V_{\rm eff}(\theta)=E_\theta$.
\label{fig-theta}}
\end{figure}

\begin{figure}
\includegraphics*[width=60mm, angle=-90]{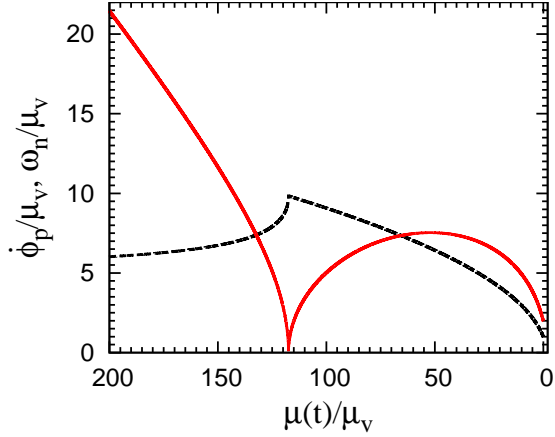}
\caption{Dimensionless precession frequency $\dot\phi_p/\mu_{\rm v}$ 
(dashed curve) and nutation frequency $\omega_n/\mu_{\rm v}$ 
(solid curve) as functions of $\mu(t)/\mu_{\rm v}$
for the neutrino gyroscope shown in Fig.~\ref{fig-theta}.\label{fig-nutpre}}
\end{figure}

We now extend the discussion in Sec.~\ref{sec-cnnu} for a constant
$\mu$ to the case where $\mu(t)$ slowly decreases with time
from a large initial value $\mu(0)\gg\mu_{\rm v}$. Specifically, we
consider the system that initially consists of monoenergetic $\nu_e$ and 
$\bar\nu_e$ only and is governed by Eqs.~(\ref{eq-s1}) and (\ref{eq-s2}).
For this system, $2\mu_{\rm v}/\delta m^2=1/9$~MeV$^{-1}$,
$\alpha=2/3$, and
\begin{equation}\label{eq-mut}
\mu(t)=\mu(0)\left[1-\sqrt{1-\left(\frac{\tau}{t+\tau}\right)^2}\,\right]^2,
\end{equation} 
where 
$\mu(0)=2.52\times 10^{5}\mu_{\rm v}$ and $\tau=8.45/\mu_{\rm v}$. 
The above form of $\mu(t)$ corresponds to $n_{\nu_e}(r)$ in the
supernova example. Taking $\tilde\theta_{\rm v}=10^{-5}$ (IH),
we numerically solve Eqs.~(\ref{eq-s1}) and (\ref{eq-s2}). 
Using the instantaneous $\mathbf{s}_1$ and $\mathbf{s}_2$ from 
the numerical results, we construct a gyroscope at each specific value
of $\mu$. We show $\cos\theta$ (dashed curve) for the series of 
gyroscopes as a function of $\mu(t)/\mu_{\rm v}$ in Fig.~\ref{fig-theta}. 
Using $\sigma$ and $Q$ for each gyroscope
and the fact that $S_z$ is conserved even when $\mu$ changes
with time [see Eq.~(\ref{eq-sg})], we construct an ``instantaneous''
$V_{\rm eff}(\theta)$ from Eq.~(\ref{eq-veff}). We define
$\theta_{\rm max}$ and $\theta_{\rm min}$ as the solutions to
$V_{\rm eff}(\theta)=E_\theta$, where $E_\theta$ is the
corresponding instantaneous effective energy [see Eq.~(\ref{eq-egy_eff})].
We show the evolution of $\cos\theta_{\rm max}$ and 
$\cos\theta_{\rm min}$ as
the dotted curves in Fig.~\ref{fig-theta}.
We also calculate $\theta_p$ corresponding to the minimum of
the instantaneous $V_{\rm eff}(\theta)$ and show the evolution of
$\cos\theta_p$ as 
the dot-dashed curve in Fig.~\ref{fig-theta}. Finally,
we calculate $\dot\phi_p$ and $\omega_n$ from
Eqs.~(\ref{eq-sz}) and (\ref{eq-d2vdt}) and show 
$\dot\phi_p/\mu_{\rm v}$ (dashed curve) 
and $\omega_n/\mu_{\rm v}$ (solid curve) as
functions of $\mu(t)/\mu_{\rm v}$ in Fig.~\ref{fig-nutpre}.

It can be seen from Fig.~\ref{fig-theta} that 
the general trend of $\theta$ follows the evolution of $\theta_p$. 
In other words, nutation of the neutrino gyroscope occurs around 
the minimum of the instantaneous $V_{\rm eff}(\theta)$ as $\mu(t)$
slowly decreases from a large initial value $\mu(0)\gg\mu_{\rm v}$.
It is also clear that the motion of the gyroscope falls into two
distinct regimes separated by a critical 
$\mu_{\rm cr}\approx 119\mu_{\rm v}$. For $\mu>\mu_{\rm cr}$,
the nutation amplitude is extremely small and 
$\theta\approx\theta_p$ to very good approximation. As $\mu$
drops below $\mu_{\rm cr}$, $\theta$ initially stays small even as
$\theta_p$ increases. This can be understood from Fig.~\ref{fig-nutpre},
which shows that the nutation frequency $\omega_n$ is small at 
$\mu\sim\mu_{\rm cr}$ and is practically zero at $\mu_{\rm cr}$.
As $\mu$ decreases further, $\omega_n$ becomes sufficiently
large and $\theta$ starts to oscillate around $\theta_p$.
The amplitude of this oscillation is also that of nutation and 
can be taken as $\eta\approx(\theta_{\rm max}-\theta_{\rm min})/2$.
The longer $\omega_n$ stays small at $\mu\sim\mu_{\rm cr}$,
the larger $\eta$ is for $\mu<\mu_{\rm cr}$.

We have constructed the series of gyroscopes for specific
values of $\mu$ using the instantaneous $\mathbf{s}_1$ and 
$\mathbf{s}_2$ numerically calculated for the $\mu(t)$ in 
Eq.~(\ref{eq-mut}). In fact, so long as $\mu(t)$ slowly decreases from 
some large initial value $\mu(0)\gg\mu_{\rm v}$, the
characteristics of such gyroscopes essentially depend on the values
of $\mu$ but not the specific functional form of $\mu(t)$. To see this,
we consider the set of parameters $\sigma$, $Q$, and $\theta_p$
that characterize the gyroscope at a specific value of $\mu$.
We can choose three equations to solve for these parameters 
as follows. From the definitions of $\sigma$,
$\mathbf{S}$, and $\mathbf{Q}$, we obtain
\begin{subequations}
\begin{align}
\sigma Q&=\frac{1-\alpha^2}{4}-\frac{\mu_{\rm v}}{\mu}S_z,
\label{eq-sigq}\\
S^2+Q^2+2\frac{\mu_{\rm v}}{\mu}Q\cos\theta+
\left(\frac{\mu_{\rm v}}{\mu}\right)^2&=\frac{1+\alpha^2}{2}.
\label{eq-qst}
\end{align}
\end{subequations}
Applying Eqs.~(\ref{eq-sz}) and (\ref{eq-qst}) to $\theta=\theta_p$, 
at which $\dot\theta$ reaches its maximum value of 
$(\dot\theta)_{\rm max}\approx\eta\omega_n$
[see Eqs.~(\ref{eq-theta}) and (\ref{eq-dottheta})], we further
obtain
\begin{subequations}
\begin{align}
\frac{\dot\phi_p}{\mu}\sin^2\theta_p+\sigma\cos\theta_p&=S_z,
\label{eq-sz2}\\
\left(\frac{\eta\omega_n}{\mu}\right)^2+
\left(\frac{\dot\phi_p}{\mu}\right)^2\sin^2\theta_p+\sigma^2
+Q^2+2\frac{\mu_{\rm v}}{\mu}Q\cos\theta_p+
\left(\frac{\mu_{\rm v}}{\mu}\right)^2&\approx\frac{1+\alpha^2}{2}.
\label{eq-qst2}
\end{align}
\end{subequations}
Treating $\eta$ as a small parameter and ignoring the $\eta^2$ term 
in Eq.~(\ref{eq-qst2}), we can solve this equation
along with Eqs.~(\ref{eq-sigq}) and (\ref{eq-sz2}) to obtain 
$\sigma$, $Q$, and $\theta_p$ [note that $\dot\phi_p$ 
is given in terms of $\sigma$, $Q$, and $\theta_p$ by 
Eq.~(\ref{eq-pre}); see Appendix~\ref{app-gyro} for a different but 
equivalent method to obtain these parameters]. 
As this approximate solution assumes $\eta=0$, it is a
``pure-precession'' solution \cite{0703776}, for which the gyroscope always
stays at the minimum of the instantaneous $V_{\rm eff}(\theta)$.

For $\tilde\theta_{\rm v}\ll 1$, the pure-precession solution gives
\begin{subequations}
\begin{align}
\theta_p&\ll 1,\\
\dot\phi_p&\to\frac{1+\alpha}{1-\alpha}\mu_{\rm v},\label{eq-wp8}\\
\omega_n&\to\frac{1-\alpha}{2}\mu-2\left(\frac{1+\alpha}{1-\alpha}\right)\mu_{\rm v},
\label{eq-wnst}
\end{align}
\end{subequations}
at $\mu\gg\mu_{\rm v}$ (see  Appendix~\ref{app-rnu} for more detailed discussion
of the initial motion of the neutrino gyroscope), and
\begin{subequations}
\begin{align}
\theta_p&\to\pi,\\
 \dot\phi_p&\to\mu_{\rm v},\label{eq-wp0}\\
\omega_n&\to 2\mu_{\rm v},
\end{align}
\end{subequations}
at $\mu\ll\mu_{\rm v}$. In general,
the $\theta_p$, $\dot\phi_p$, and $\omega_n$
calculated for the pure-precession solution are within $\sim1\%$ of the
values shown in Figs.~\ref{fig-theta} and \ref{fig-nutpre}. 

Equation~(\ref{eq-wnst}) suggests that $\omega_n$ becomes very small
as $\mu$ decreases to some critical value $\mu_{\rm cr}$. 
In Appendix~\ref{app-critical}, we show that
\begin{equation}
\mu_{\rm cr}=\frac{4\mu_{\rm v}}{\left( 1-\sqrt{\alpha} \right)^2},
\end{equation}
which is $\mu_{\rm cr}\approx119\mu_{\rm v}$ for $\alpha=2/3$, in excellent
agreement with Figs.~\ref{fig-theta} and \ref{fig-nutpre}.
At $\mu=\mu_{\rm cr}$, we have (see Appendix~\ref{app-critical})
\begin{subequations}
\begin{align}
\theta_{p,{\rm cr}}&\approx\frac{2\sqrt{2}\alpha^{1/6}}
{(1+\sqrt{\alpha})^{2/3}}\tilde\theta_{\rm v}^{2/3},\\
\dot\phi_{p,{\rm cr}}&\approx\frac{\mu_{\rm cr}S_z}{2}
\approx\frac{1+\sqrt{\alpha}}{1-\sqrt{\alpha}}\mu_{\rm v},\label{eq-wpc}\\
\omega_{n,{\rm cr}}&\approx\frac{\sqrt{3}}{2}\mu_{\rm cr}S_z\theta_{p,{\rm cr}}
\approx\frac{2\sqrt{6}\alpha^{1/6}(1+\sqrt{\alpha})^{1/3}}
{1-\sqrt{\alpha}}\tilde\theta_{\rm v}^{2/3}\mu_{\rm v}.\label{eq-wnc}
\end{align}
\end{subequations}
As $\theta_p\ll 1$ at $\mu\gg\mu_{\rm v}$ and at $\mu=\mu_{\rm cr}$
for $\tilde\theta_{\rm v}\ll 1$, the neutrino gyroscope stays in the upright 
position and behaves like a sleeping top (e.g., \cite{0608695,0703776})
at $\mu\geq\mu_{\rm cr}$ (see Fig.~\ref{fig-theta}).

\section{Resonances Driven by The Neutrino Gyroscope}\label{sec-res}
In this section we return to the system initially consisting of $\nu_e$ and 
$\bar\nu_e$ with spectra of the form in Eq.~(\ref{eq-normspec}).
As discussed in Secs.~\ref{sec-nueanue} and \ref{sec-gyromf}, 
the approximate mean field of NFIS's for this system can be
described by the neutrino gyroscope. We now try to understand the
evolution of an individual NFIS in the system in terms of its response
to the neutrino gyroscope. 
Specifically, we study the evolution of
$\mathbf{s}_\omega$ governed by
\begin{equation}
\label{eq-smf1}
\frac{d}{dt}\mathbf{s}_\omega  =\mathbf{s}_\omega\times
[\omega\mathbf{H}_{\rm v}-\mu(t)\mathbf{S}]
=[\omega\mathbf{\hat e}_z^{\rm I}+\mu(t)\mathbf{S}]\times\mathbf{s}_\omega,
\end{equation}
where $\mathbf{S}$ is the total angular momentum of the neutrino gyroscope
discussed in Sec.~\ref{sec-vnnu}.

\subsection{Precession-Driven Resonance\label{sec-respre}}
\begin{figure}
\includegraphics*[width=80mm]{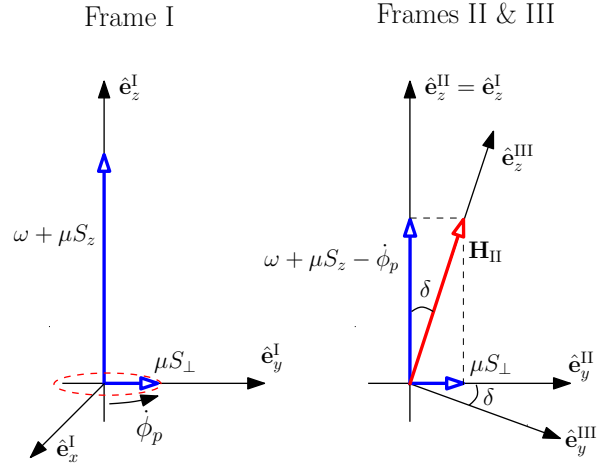}
\caption{Illustration of relations among Frames I, II, and III. Frame I is fixed
to the laboratory. Frame II has the same $z$-axis as Frame I and rotates
around this axis with an angular velocity $\dot\phi_p\mathbf{\hat e}_z^{\rm I}$
relative to Frame I. When only precession of the neutrino gyroscope is considered,
the net effective field interacting with an individual NFIS $\mathbf{s}_\omega$
has two components in Frame I: one fixed in the direction of 
$\mathbf{\hat e}_z^{\rm I}$ and the other rotating with an angular velocity 
$\dot\phi_p\mathbf{\hat e}_z^{\rm I}$. It is convenient to use Frame II 
to discuss the precession-driven resonance as the net effective field becomes
a non-rotating vector $\mathbf{H}_{\rm II}$ in this frame.
Frame III has the same $x$-axis as Frame II but its $z$-axis is in the direction
of $\mathbf{H}_{\rm II}$, which makes an angle $\delta$ with respect to
$\mathbf{\hat e}_z^{\rm I}$. Frame III is used in
discussing the nutation-driven resonance. See text for details.
\label{fig-frames}}
\end{figure}

\begin{figure}
\includegraphics*[width=60mm, angle=-90]{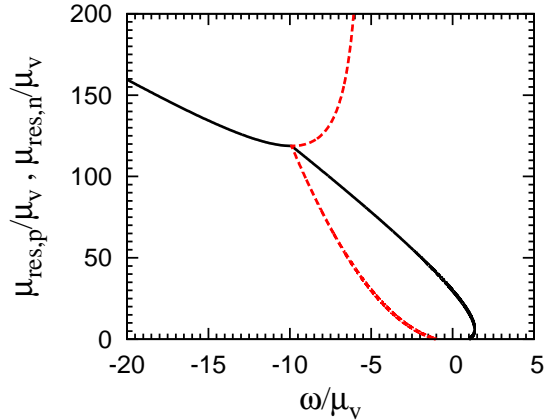}
\caption{Values of $\mu_{{\rm res},p}/\mu_{\rm v}$ (solid curve)
and $\mu_{{\rm res},n}/\mu_{\rm v}$ (dashed curve) as functions of 
$\omega/\mu_{\rm v}$. An NFIS $\mathbf{s}_\omega$ goes through
the precession-driven (nutation-driven) resonance at
$\mu=\mu_{{\rm res},p}$ ($\mu_{{\rm res},n}$) as the neutrino
gyroscope in Fig.~\ref{fig-theta} evolves through $\mu(t)$.\label{fig-wres}}
\end{figure}

We first ignore nutation and consider only precession of the neutrino gyroscope. 
With $\eta=0$ (and $\gamma=0$ as noted in Sec.~\ref{sec-cnnu}), 
Eq.~(\ref{eq-sxy}) becomes
\begin{equation}\label{eq-sxyp}
S_x+iS_y\approx iS_\perp\exp(i\dot\phi_pt),
\end{equation}
where
\begin{equation}\label{eq-sperp}
S_\perp=\left(\frac{\dot\phi_p}{\mu}\cos\theta_p-\sigma\right)\sin\theta_p.
\end{equation}
Equation~(\ref{eq-sxyp}) represents a vector rotating
in the $xy$-plane of Frame I. Let Frame II rotate with an 
angular velocity $\dot\phi_p\mathbf{\hat e}_z^{\rm I}$
relative to Frame I. Then $\mathbf{S}$ is a 
non-rotating vector in Frame II and can be chosen as
\begin{equation}\label{eq-sii}
\mathbf{S}\approx S_z\mathbf{\hat e}_z^{\rm I}
+S_\perp\mathbf{\hat e}_y^{\rm II},
\end{equation}
where $\mathbf{\hat e}_y^{\rm II}$ is the unit vector in
the $y$-direction of Frame II (see Fig.~\ref{fig-frames}). 
We rewrite Eq.~(\ref{eq-smf1}) in this frame as
\begin{equation}
\label{eq-smfp}
\frac{d}{dt}\mathbf{s}_\omega=
[(\omega+\mu S_z-\dot\phi_p)
\mathbf{\hat e}_z^{\rm I}
+\mu S_\perp\mathbf{\hat e}_y^{\rm II}]
\times\mathbf{s}_\omega\equiv\mathbf{H}_{\rm II}
\times\mathbf{s}_\omega.
\end{equation}
Note that in the above equation $\omega$ and $S_z$
are constants but $\dot\phi_p$ and $S_\perp$ are 
functions of $\mu(t)$. A resonance occurs when the $z$-component 
of $\mathbf{H}_{\rm II}$ vanishes.
We refer to this as the precession-driven resonance.
We denote the value of $\mu$ at which an individual NFIS
$\mathbf{s}_\omega$ goes through this resonance as
$\mu_{{\rm res},p}$, which can be obtained from
\begin{equation}\label{eq-resp}
\omega=\dot\phi_p(\mu_{{\rm res},p})-\mu_{{\rm res},p}S_z.
\end{equation}
Here $\dot\phi_p(\mu_{{\rm res},p})$ is the value of
$\dot\phi_p$ at $\mu=\mu_{{\rm res},p}$.
Using the neutrino gyroscope in Fig.~\ref{fig-theta},
we show $\mu_{{\rm res},p}/\mu_{\rm v}$ 
as a function of $\omega/\mu_{\rm v}$ 
(solid curve) in Fig.~\ref{fig-wres} (note that the pure-precession
solution gives essentially the same result).

\begin{figure}
\includegraphics*[width=120mm]{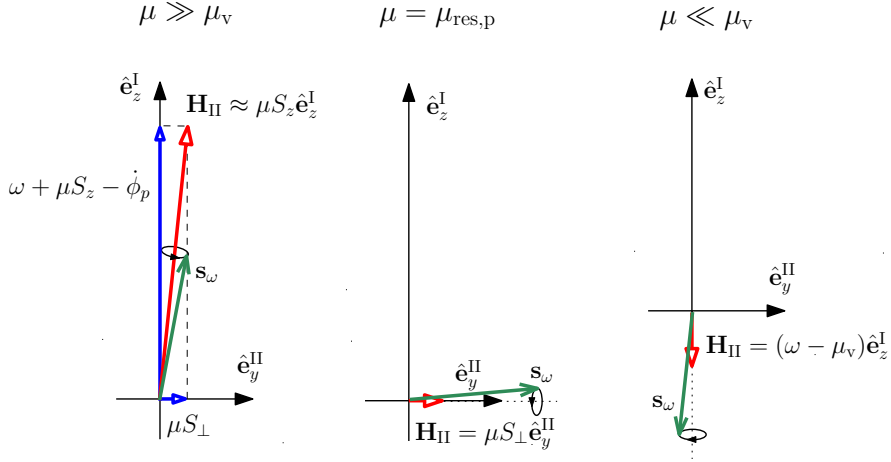}
\caption{Illustration of nearly full conversion of an initial $\nu_e$
represented by $\mathbf{s}_\omega$ with $\omega<\mu_{\rm v}$
following adiabatic evolution through a precession-driven 
resonance. In Frame II, adiabatic evolution corresponds to
precession of $\mathbf{s}_\omega$ around $\mathbf{H}_{\rm II}$
with a fixed angle between the two vectors. At $\mu\gg\mu_{\rm v}$,
$\mathbf{H}_{\rm II}$ nearly coincides with $\mathbf{\hat e}_z^{\rm I}$.
At $\mu=\mu_{{\rm res},p}$ corresponding to the resonance, the
$z$-component of $\mathbf{H}_{\rm II}$ vanishes. At $\mu\ll\mu_{\rm v}$,
$\mathbf{H}_{\rm II}$ is opposite to $\mathbf{\hat e}_z^{\rm I}$ for
$\omega<\mu_{\rm v}$. Consequently, the initial and final directions of 
$\mathbf{H}_{\rm II}$ are nearly opposite, and the same is true of 
$\mathbf{s}_\omega$.\label{fig-resp}}
\end{figure}

The evolution of $\mathbf{s}_\omega$ in 
Frame II based on Eq.~(\ref{eq-smfp}) is very similar to the 
MSW effect (see Fig.~\ref{fig-resp}). 
At a specific time $t$, $\mathbf{s}_\omega$ precesses 
around the instantaneous $\mathbf{H}_{\rm II}$ with an angular
velocity $\mathbf{H}_{\rm II}$. For slowly varying
$\mu(t)$, the evolution of
$\mathbf{s}_\omega$ is adiabatic in that the precession
adjusts to the instantaneous angular velocity (the
direction and magnitude of which are both changing slowly
in general) but the
angle between $\mathbf{s}_\omega$ and $\mathbf{H}_{\rm II}$
remains fixed. Therefore, the initial $\nu_e$ 
or $\bar\nu_e$ represented by $\mathbf{s}_\omega$ remains 
in the same flavor following adiabatic evolution if the initial and final 
directions of $\mathbf{H}_{\rm II}$ are the same,
but is fully converted into a $\nu_x$ or $\bar\nu_x$ 
if the initial and final directions of $\mathbf{H}_{\rm II}$ are opposite.
For the neutrino gyroscope under consideration, 
$\mathbf{H}_{\rm II}\approx \mu S_z\mathbf{\hat e}_z^{\rm I}$ 
at $t=0$ corresponding to $\mu\gg\mu_{\rm v}$, where we have used 
$\mu S_z\gg|\omega|,\dot\phi_p,\mu|S_\perp|$ 
in this limit. At large times corresponding to $\mu\ll\mu_{\rm v}$,
$\mathbf{H}_{\rm II}\approx(\omega-\mu_{\rm v})\mathbf{\hat e}_z^{\rm I}$, 
where we have used $\dot\phi_p=\mu_{\rm v}$ in this limit.
Consequently, the initial and final directions of $\mathbf{H}_{\rm II}$ are
the same for $\omega>\mu_{\rm v}$ but are opposite
for $\omega<\mu_{\rm v}$. In the latter case, a precession-driven 
resonance occurs when the $z$-component of 
$\mathbf{H}_{\rm II}$ vanishes before
the direction of $\mathbf{H}_{\rm II}$ is reversed (see Fig.~\ref{fig-resp}). 
Thus, when only precession-driven 
resonance matters and adiabatic evolution applies,
an initial $\nu_e$ with $\omega>\mu_{\rm v}$ remains as a
$\nu_e$, while an initial $\nu_e$ with $\omega<\mu_{\rm v}$ or 
an initial $\bar\nu_e$ (with $\omega<0$) is fully converted
into a $\nu_x$ or $\bar\nu_x$, respectively. This is basically the
explanation for the stepwise spectral swap originally discovered in 
\cite{0606616} (see also discussion in \cite{0705.1830}).

\begin{figure}
\includegraphics*[width=60mm, angle=-90]{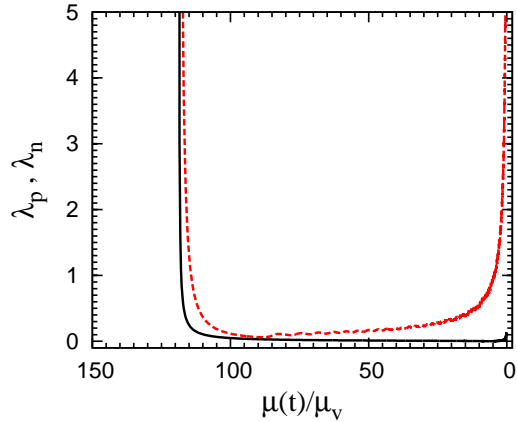}
\caption{Adiabaticity parameters $\lambda_p$ (solid curve) and 
$\lambda_n$ (dashed curve) for evolution through precession-driven and
nutation-driven resonances, respectively, as functions
of $\mu(t)/\mu_{\rm v}$ for the neutrino gyroscope shown
in Fig.~\ref{fig-theta}.\label{fig-lampn}}
\end{figure}

For adiabatic evolution, the rate at which the direction of
$\mathbf{H}_{\rm II}$ changes must be slow compared with
the precession frequency of $\mathbf{s}_\omega$:
\begin{equation}\label{eq-adp}
\left|\frac{d}{dt}\frac{\mathbf{H}_{\rm II}}{|\mathbf{H}_{\rm II}|}\right|
=\frac{|\mathbf{H}_{\rm II}\times d\mathbf{H}_{\rm II}/dt|}
{|\mathbf{H}_{\rm II}|^2}\ll |\mathbf{H}_{\rm II}|.
\end{equation}
The above condition is most stringent at resonance when the
$z$-component of $\mathbf{H}_{\rm II}$ vanishes and 
$|\mathbf{H}_{\rm II}|$ becomes very small (see Fig.~\ref{fig-resp}).
We define the adiabaticity parameter for this precession-driven
resonance as
\begin{equation}
\lambda_p\equiv\frac{|\mathbf{H}_{\rm II}\times 
d\mathbf{H}_{\rm II}/dt|_{\rm res}}
{|\mathbf{H}_{\rm II}|_{\rm res}^3}=
\frac{|d(\dot\phi_p-\mu S_z)/dt|}
{(\mu S_\perp)^2}.
\end{equation}
Adiabatic evolution obtains for $\lambda_p\ll1$ (note that $\lambda_p$
is defined differently from the usual adiabaticity parameter
for the conventional MSW effect). 
Using the neutrino gyroscope in Fig.~\ref{fig-theta},
we show $\lambda_p$ as a function of 
$\mu(t)/\mu_{\rm v}$ (solid curve) in Fig.~\ref{fig-lampn}. 
It can be seen from this figure that evolution through the precession-driven
resonance is adiabatic at $\mu<115\mu_{\rm v}$, 
but is extremely non-adiabatic at $\mu\gtrsim\mu_{\rm cr}\approx 119\mu_{\rm v}$. 
As $\mu\gtrsim\mu_{\rm cr}$ corresponds
to the sleeping-top regime with $\theta_p\ll 1$, the small values of
$S_\perp\propto\sin\theta_p$ [see Eq.~(\ref{eq-sperp})] result in 
$\lambda_p\gg 1$ in this regime.

\subsection{Nutation-Driven Resonance\label{sec-resnut}}
Now we consider both precession and nutation of the neutrino
gyroscope. The terms proportional to $\eta$ in Eq.~(\ref{eq-sxy})
contain the factors $\exp(i\omega_nt)$ and
$\exp(-i\omega_nt)$, which correspond to
rotation with angular velocities 
$\omega_n\mathbf{\hat e}_z^{\rm I}$ and
$-\omega_n\mathbf{\hat e}_z^{\rm I}$, respectively, 
relative to Frame II. However,
frames rotating with these angular velocities are not
convenient to use because 
$\mu(t)S_\perp\mathbf{\hat e}_y^{\rm II}$, and hence 
$\mathbf{H}_{\rm II}$, rotate in such frames. To find
the appropriate frames, we first consider Frame III with
its axes defined by the unit vectors (see Fig.~\ref{fig-frames})
\begin{subequations}
\begin{align}
\mathbf{\hat e}_x^{\rm III}&=\mathbf{\hat e}_x^{\rm II},\\
\mathbf{\hat e}_y^{\rm III}&=\cos\delta\mathbf{\hat e}_y^{\rm II}
-\sin\delta\mathbf{\hat e}_z^{\rm I},\\
\mathbf{\hat e}_z^{\rm III}&=\mathbf{H}_{\rm II}/|\mathbf{H}_{\rm II}|
=\sin\delta\mathbf{\hat e}_y^{\rm II}+\cos\delta\mathbf{\hat e}_z^{\rm I},
\label{eq-coord3z}
\end{align}
\end{subequations}
where
\begin{subequations}
\begin{align}
\cos\delta&=\frac{\omega+\mu S_z-\dot\phi_p}
{|\mathbf{H}_{\rm II}|},\label{eq-cosdelta}\\
\sin\delta&=\frac{\mu S_\perp}{|\mathbf{H}_{\rm II}|}.
\end{align}
\end{subequations}
Note that just like Frame II, Frame III also rotates with 
an angular velocity 
$\dot\phi_p\mathbf{\hat e}_z^{\rm I}$ relative to Frame I.
Using Eq.~(\ref{eq-sxy}) (with $\beta=\gamma=0$ as noted in
Sec.~\ref{sec-cnnu}), we write $\mathbf{S}$ in Frame III as
\begin{align}\label{eq-s3}
\mathbf{S}&\approx S_z\mathbf{\hat e}_z^{\rm I}+
S_\perp\mathbf{\hat e}_y^{\rm II}+\eta\frac{\dot\phi_p}{\mu}\sin\delta
\cos(\omega_nt)\mathbf{\hat e}_z^{\rm III}\nonumber\\
&+\eta\frac{\dot\phi_p}{\mu}\frac{2\dot\phi_p-\mu S_z}{\omega_n}
\sin(\omega_nt)\mathbf{\hat e}_x^{\rm III}+
\eta\frac{\dot\phi_p}{\mu}\cos\delta
\cos(\omega_nt)\mathbf{\hat e}_y^{\rm III}.
\end{align}
The last two terms in the above expression can be rewritten as
two vectors rotating with angular velocities
$\omega_n\mathbf{\hat e}_z^{\rm III}$ and
$-\omega_n\mathbf{\hat e}_z^{\rm III}$, respectively,
relative to Frame III:
\begin{equation}\label{eq-sxy3}
(S_x+iS'_y)_{\rm III}\equiv i\frac{\eta}{2}\frac{\dot\phi_p}{\mu}
\left[\left(\cos\delta-\frac{2\dot\phi_p-\mu S_z}{\omega_n}\right)
e^{i\omega_nt}
+\left(\cos\delta+\frac{2\dot\phi_p-\mu S_z}{\omega_n}\right)
e^{-i\omega_nt}\right].
\end{equation}

As we will see shortly,
a new resonance occurs for $|\mathbf{H}_{\rm II}|=\omega_n$.
Using Eqs.~(\ref{eq-d2vdt}),
(\ref{eq-pre}), (\ref{eq-sz2}), and (\ref{eq-smfp}),
we can rewrite the above resonance condition as:
\begin{equation}
(\omega+\dot\phi_p)(\omega+2\mu S_z-3\dot\phi_p)=0.
\end{equation}
The term with the factor $\exp(-i\omega_nt)$ in Eq.~(\ref{eq-sxy3})
vanishes for $\omega=-\dot\phi_p$, while that with the factor
$\exp(i\omega_nt)$ vanishes for $\omega=3\dot\phi_p-2\mu S_z$. 
We will see that the new resonance corresponds to $\omega=-\dot\phi_p$.
So we can ignore the term with the factor $\exp(-i\omega_nt)$ in 
Eq.~(\ref{eq-sxy3}) when treating this resonance.
We choose Frame IV to rotate with an angular velocity
$\omega_n\mathbf{\hat e}_z^{\rm III}$ relative to Frame III.
The term with the factor $\exp(i\omega_nt)$
in Eq.~(\ref{eq-sxy3}) represents a vector parallel to the unit vector
$\mathbf{\hat e}_y^{\rm IV}$ in the $y$-direction of Frame IV. In this
frame Eq.~(\ref{eq-smf1}) effectively becomes
\begin{equation}\label{eq-smf3pl}
\frac{d}{dt}\mathbf{s}_\omega=\left[\left(|\mathbf{H}_{\rm II}|-
\omega_n+\eta\dot\phi_p\sin\delta\cos\omega_nt\right)
\mathbf{\hat e}_z^{\rm III}+\frac{\eta\dot\phi_p}{2}
\left(\cos\delta-\frac{2\dot\phi_p-\mu S_z}{\omega_n}\right)
\mathbf{\hat e}_y^{\rm IV}\right]\times\mathbf{s}_\omega.
\end{equation}
It can be seen that a new resonance indeed occurs when
$|\mathbf{H}_{\rm II}|=\omega_n$ if we ignore the small contribution
proportional to $\eta$ in the term associated with 
$\mathbf{\hat e}_z^{\rm III}$ in the above equation.
We refer to this as the nutation-driven resonance because it is driven
by the nutation-dependent component of $\mathbf{S}$.
We denote the value of $\mu$ at which an individual NFIS
$\mathbf{s}_\omega$ goes through this resonance as
$\mu_{{\rm res},n}$, which can be obtained from
\begin{equation}
\omega=-\dot\phi_p(\mu_{{\rm res},n}).
\end{equation}
Here $\dot\phi_p(\mu_{{\rm res},n})$ is the value of $\dot\phi_p$
at $\mu=\mu_{{\rm res},n}$.
Using the neutrino gyroscope in Fig.~\ref{fig-theta},
we show $\mu_{{\rm res},n}/\mu_{\rm v}$ 
as a function of $\omega/\mu_{\rm v}$ 
(dashed curve) in Fig.~\ref{fig-wres} (note again that the pure-precession
solution gives essentially the same result).

To see that $\omega=3\dot\phi_p-2\mu S_z$, which also gives
$|\mathbf{H}_{\rm II}|=\omega_n$, does not correspond to a resonance,
we recall that the term with the factor $\exp(i\omega_nt)$ in 
Eq.~(\ref{eq-sxy3}) vanishes for this $\omega$.
We choose Frame V to rotate with an angular velocity
$-\omega_n\mathbf{\hat e}_z^{\rm III}$ relative to Frame III and
rewrite Eq.~(\ref{eq-smf1}) in Frame V effectively as
\begin{equation}\label{eq-smf5}
\frac{d}{dt}\mathbf{s}_\omega=\left[\left(|\mathbf{H}_{\rm II}|+
\omega_n+\eta\dot\phi_p\sin\delta\cos\omega_nt\right)
\mathbf{\hat e}_z^{\rm III}+\frac{\eta\dot\phi_p}{2}
\left(\cos\delta+\frac{2\dot\phi_p-\mu S_z}{\omega_n}\right)
\mathbf{\hat e}_y^{\rm V}\right]\times\mathbf{s}_\omega,
\end{equation}
where $\mathbf{\hat e}_y^{\rm V}$ is the unit vector
in the $y$-direction of Frame V. It can be seen that the term
associated with $\mathbf{\hat e}_z^{\rm III}$ in the above equation
never vanishes, and consequently, there is no resonance
for $\omega=3\dot\phi_p-2\mu S_z$.

The adiabaticity for evolution through the nutation-driven
resonance can be discussed similarly to the case of
precession-driven resonance studied in Sec.~\ref{sec-respre}.
The evolution of 
$\mathbf{s}_\omega$ in Frame IV is governed by
\begin{equation}
\frac{d}{dt}\mathbf{s}_\omega=\mathbf{H}_{\rm IV}\times
\mathbf{s}_\omega.
\end{equation}
At resonance $\mathbf{H}_{\rm IV}$ is given by
\begin{equation}\label{eq-h4res}
(\mathbf{H}_{\rm IV})_{\rm res}\approx\eta\dot\phi_p
\left(\frac{\mu S_z-2\dot\phi_p}{\omega_n}\right)
\mathbf{\hat e}_y^{\rm IV}.
\end{equation}
We define the adiabaticity parameter for the nutation-driven 
resonance as
\begin{equation}\label{eq-adn}
\lambda_n\equiv\frac{|\mathbf{H}_{\rm IV}\times 
d\mathbf{H}_{\rm IV}/dt|_{\rm res}}
{|\mathbf{H}_{\rm IV}|_{\rm res}^3}\approx
\frac{|d(|\mathbf{H}_{\rm II}|-\omega_n)/dt|_{\rm res}}
{|\mathbf{H}_{\rm IV}|_{\rm res}^2}.
\end{equation}
In Eqs.(\ref{eq-h4res}) and (\ref{eq-adn}),
we have neglected oscillatory terms proportional to
$\eta$. Using the neutrino gyroscope in Fig.~\ref{fig-theta},
we show $\lambda_n$ as a function of $\mu/\mu_{\rm v}$ 
(dashed curve) in Fig.~\ref{fig-lampn}. It can be seen from
this figure that evolution through the nutation-driven
resonance is adiabatic at $\mu/\mu_{\rm v}\sim10$--110
and becomes non-adiabatic outside this range. In particular,
evolution is extremely non-adiabatic at 
$\mu\gtrsim\mu_{\rm cr}\approx119\mu_{\rm v}$
and at $\mu\lesssim5\mu_{\rm v}$ as the
very small nutation amplitude $\eta$
(see Fig.~\ref{fig-theta}) results in $\lambda_n\gg 1$
in these two regimes.

\subsection{Evolution through Resonances Driven by the
Neutrino Gyroscope}\label{sec-evolres}
\begin{figure*}
\includegraphics*[width=16cm, angle=-90]{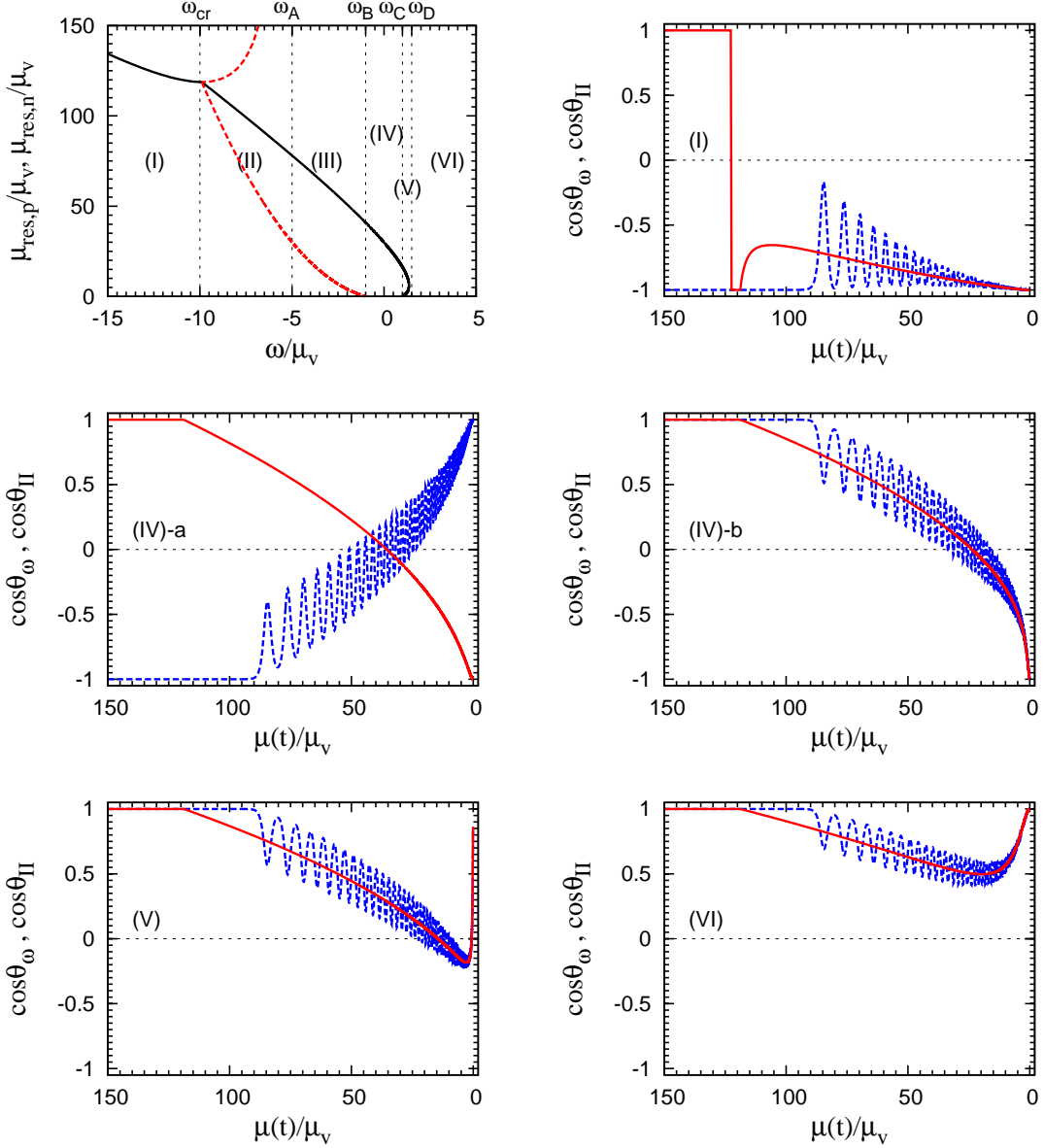}
\caption{Ranges I--VI of $\omega$ based on occurrences of
precession-driven (solid curve) and nutation-driven (dashed curve)
resonances (top left panel)
and example evolution of $\cos\theta_\omega$ (dashed curves) and 
$\cos\theta_{\rm II}$ (solid curves) as functions of $\mu(t)/\mu_{\rm v}$ 
for $\omega/\mu_{\rm v}=-12$ (I), $-0.5$ (IV-a), 0.5 (IV-b), 1.1 (V), and 
3 (VI), respectively. See text for details.\label{fig-evolres}}
\end{figure*}

Based on the discussion in Secs.~\ref{sec-respre} and \ref{sec-resnut},
an NFIS may experience two types of resonances driven by
precession and nutation of the neutrino gyroscope, respectively.
For the NFIS $\mathbf{s}_\omega$, a precession-driven resonance occurs 
at $\omega=\dot{\phi}_p(\mu_{{\rm res},p})-\mu_{{\rm res},p}S_z$, and
a nutation-driven resonance occurs at $\omega=-\dot{\phi}_p(\mu_{{\rm res},n})$.
Formally these two resonances coincide at 
$\mu_{{\rm res},p}=\mu_{{\rm res},n}=\mu_{\rm cr}$ for
$\omega=\omega_{\rm cr}\equiv -\dot{\phi}_{p,{\rm cr}}\approx
(1+\sqrt{\alpha})\mu_{\rm v}/(1-\sqrt{\alpha})$ [see Eq.~(\ref{eq-wpc})].
Noting that $\dot{\phi}_p\to(1+\alpha)\mu_{\rm v}/(1-\alpha)$ at $\mu\gg\mu_{\rm cr}$
[see Eq.~(\ref{eq-wp8})] and $\dot{\phi}_p\to\mu_{\rm v}$ at $\mu\ll\mu_{\rm v}$
[see Eq.~(\ref{eq-wp0})], we introduce 
$\omega_A\equiv-(1+\alpha)\mu_{\rm v}/(1-\alpha)$,
$\omega_B\equiv-\mu_{\rm v}$, and $\omega_C\equiv\mu_{\rm v}$ to
define ranges of $\omega$ with different resonances.
It turns out that for $\omega=\omega_D$ slightly larger than $\omega_C$,
there are two possible values for $\mu_{{\rm res},p}$. Altogether, the varieties of
resonances experienced by $\mathbf{s}_\omega$ can be classified into six
categories:
\newcounter{bean}
\begin{list}
{\Roman{bean}.}{\usecounter{bean}}
\item for $\omega<\omega_{\rm cr}$, $\mathbf{s}_\omega$ experiences only
a precession-driven resonance at $\mu_{{\rm res},p}>\mu_{\rm cr}$;
\item for $\omega_{\rm cr}<\omega<\omega_A$, $\mathbf{s}_\omega$ experiences
a nutation-driven resonance at $\mu_{{\rm res},n}>\mu_{\rm cr}$, then a 
precession-driven resonance at $\mu_{{\rm res},p}<\mu_{\rm cr}$, and finally
a second nutation-driven resonance at $\mu_{{\rm res},n}'<\mu_{{\rm res},p}$;
\item for $\omega_A<\omega<\omega_B$, $\mathbf{s}_\omega$ experiences
a precession-driven resonance at $\mu_{{\rm res},p}<\mu_{\rm cr}$ followed by
a nutation-driven resonance at $\mu_{{\rm res},n}<\mu_{{\rm res},p}$;
\item for $\omega_B<\omega<\omega_C$, $\mathbf{s}_\omega$ experiences
only a precession-driven resonance at $\mu_{{\rm res},p}<\mu_{\rm cr}$;
\item for $\omega_C<\omega<\omega_D$, $\mathbf{s}_\omega$ experiences
two precession-driven resonances at $\mu_{{\rm res},p}<\mu_{\rm cr}$ and
$\mu_{{\rm res},p}'<\mu_{{\rm res},p}$, respectively.
\item for $\omega>\omega_D$, $\mathbf{s}_\omega$ does not experience
any resonance.
\end{list}
For the neutrino gyroscope in Fig.~\ref{fig-theta}, 
$\mu_{\rm cr}\approx 119\mu_{\rm v}$,
$\omega_{\rm cr}\approx-10\mu_{\rm v}$,
$\omega_A\approx-5\mu_{\rm v}$, and
$\omega_D\approx1.5\mu_{\rm v}$. The six categories of resonances for
this example are shown in the top left panel of Fig.~\ref{fig-evolres}.

Only precession-driven resonances are involved for $\omega$ in 
ranges I, IV, and V, and there are no resonances for $\omega$ in
range VI. We first consider the overall evolution of $\mathbf{s}_\omega$
for $\omega$ in these ranges using the neutrino gyroscope in 
Fig.~\ref{fig-theta}. As nutation is unimportant for these cases, we 
focus on $\mathbf{H}_{\rm II}$ as the net effective field interacting with 
$\mathbf{s}_\omega$ [see Eq.~(\ref{eq-smfp})].
At $\mu=\mu(0)\gg\mu_{\rm cr}$, $\mathbf{H}_{\rm II}$ is essentially
in the direction of $\mathbf{\hat{e}}_z^{\rm I}$ and $\mathbf{s}_\omega$
is either aligned (initial $\nu_e$, $\omega>0$) or anti-aligned 
(initial $\bar\nu_e$, $\omega<0$) with $\mathbf{H}_{\rm II}$.
We show the subsequent evolution of $\mathbf{s}_\omega$ relative to
$\mathbf{H}_{\rm II}$ by comparing $\cos\theta_\omega\equiv
\mathbf{s}_\omega\cdot\mathbf{\hat{e}}_z^{\rm I}/|\mathbf{s}_\omega|$ 
(dashed curve) with $\cos\theta_{\rm II}\equiv\mathbf{H}_{\rm II}\cdot
\mathbf{\hat{e}}_z^{\rm I}/|\mathbf{H}_{\rm II}|$ (solid curve) in 
Fig.~\ref{fig-evolres}. For $\omega/\mu_{\rm v}=-12$ in range I,
$\mathbf{s}_\omega$ is initially anti-aligned with $\mathbf{H}_{\rm II}$
($\cos\theta_\omega\approx-1$ but $\cos\theta_{\rm II}\approx 1$).
At $\mu/\mu_{\rm v}\approx 122$, $\cos\theta_{\rm II}$ vanishes
and a resonance occurs. However, evolution through this 
resonance is extremely non-adiabatic (see Fig.~\ref{fig-lampn}).
Consequently, $\cos\theta_\omega$ is unaffected while 
$\cos\theta_{\rm II}$ changes drastically from $\approx 1$ to $\approx -1$ 
immediately after the resonance. Subsequent evolution of 
$\mathbf{s}_\omega$ is essentially adiabatic with
$\cos\theta_\omega$ oscillating around $\cos\theta_{\rm II}$ and
eventually settling to $\approx -1$ again. This kind of evolution applies
to all $\omega$ in range I, for which there is no net flavor transformation.
For $\omega/\mu_{\rm v}=-0.5$ (0.5) in range IV, there is a resonance
at $\mu/\mu_{\rm v}\approx 36$ (24). Evolution through
the resonance is adiabatic (see Fig.~\ref{fig-lampn}) and 
$\mathbf{s}_\omega$ stays anti-aligned (aligned) with 
$\mathbf{H}_{\rm II}$ during the entire evolution. As the initial and final 
directions of $\mathbf{H}_{\rm II}$ are opposite, there is full flavor
conversion for $\omega$ in range IV. For $\omega/\mu_{\rm v}=1.1$
in range V, there are two resonances at $\mu/\mu_{\rm v}\approx 14$ 
and 1, respectively. Evolution through both resonances is essentially
adiabatic (see Fig.~\ref{fig-lampn}) and $\mathbf{s}_\omega$ stays
aligned with $\mathbf{H}_{\rm II}$ during the entire evolution. 
As $\cos\theta_{\rm II}$ changes sign twice, the initial and final 
directions of $\mathbf{H}_{\rm II}$ are the same and there is no net
flavor transformation for $\omega$ in range V. Finally,
for $\omega/\mu_{\rm v}=3$ in range VI, there is no 
resonance and evolution is adiabatic. So $\cos\theta_\omega$
oscillates around $\cos\theta_{\rm II}$, indicating that 
$\mathbf{s}_\omega$ is always aligned with $\mathbf{H}_{\rm II}$.
There is no net flavor transformation for $\omega$ in range VI.

\begin{figure*}
\includegraphics*[width=10cm, angle=-90]{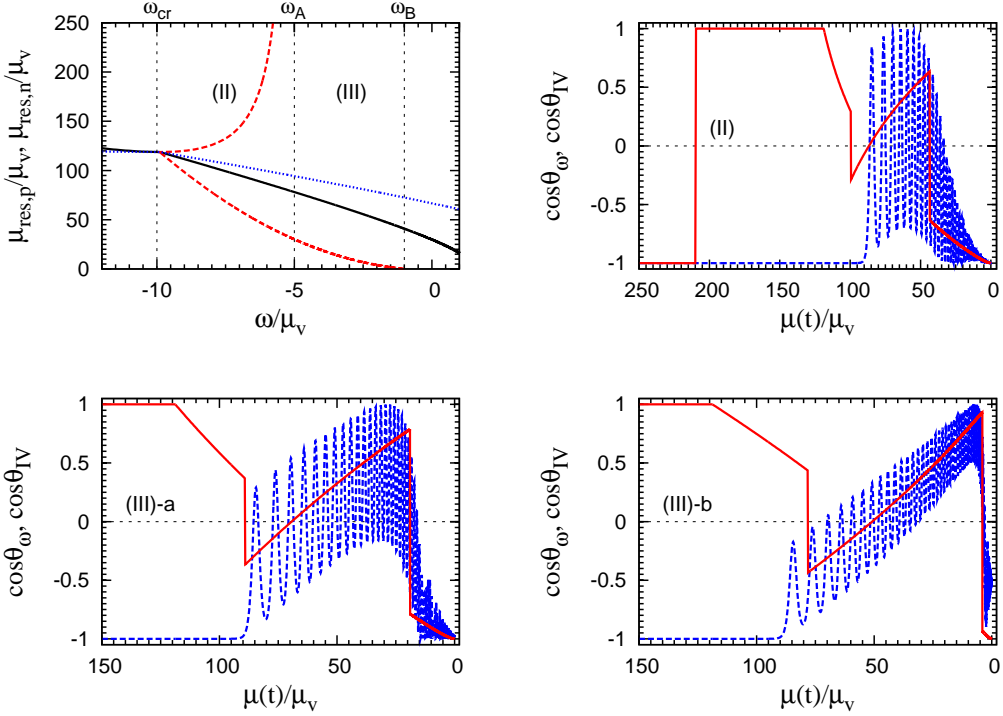}
\caption{Ranges II and III of $\omega$ based on occurrences of
precession-driven (solid curve) and nutation-driven (dashed curve)
resonances (top left panel)
and example evolution of $\cos\theta_\omega$ (dashed curves) and 
$\cos\theta_{\rm IV}$ (solid curves) as functions of $\mu(t)/\mu_{\rm v}$ 
for $\omega/\mu_{\rm v}=-6$ (II), $-4$ (III-a), and $-2$ (III-b), 
respectively. The dotted curve in the top left panel shows the 
value of $\mu/\mu_{\rm v}$ for which $\omega=3\dot\phi_p-2\mu S_z$
as a function of $\omega/\mu_{\rm v}$. Note that no resonance
occurs at this $\mu$ although the corresponding $\cos\theta_{\rm IV}$ 
vanishes. See text for details.\label{fig-evolnp}}
\end{figure*}

Resonances driven by both precession and nutation of the neutrino
gyroscope are involved for $\omega$ in ranges II and III. We discuss
the evolution of $\mathbf{s}_\omega$ for these ranges using
$\mathbf{H}_{\rm IV}$ as the net effective field. We define
$\cos\theta_{\rm IV}\equiv\mathbf{H}_{\rm IV}\cdot
\mathbf{\hat{e}}_z^{\rm I}/|\mathbf{H}_{\rm IV}|$.
Neglecting terms proportional to $\eta$, we obtain
$\cos\theta_{\rm IV}\approx{\rm sgn}(|\mathbf{H}_{\rm II}|-\omega_n)
\cos\theta_{\rm II}$ (see Sec.~\ref{sec-resnut}), where sgn$(x)$ is the
sign of $x$. Using the neutrino gyroscope in Fig.~\ref{fig-theta},
we compare the evolution of $\cos\theta_\omega$ and
$\cos\theta_{\rm IV}$ for $\omega$ in ranges II and III in Fig.~\ref{fig-evolnp}.
For $\omega/\mu_{\rm v}=-6$ in range II, $\mathbf{s}_\omega$
is initially aligned with $\mathbf{H}_{\rm IV}$. A nutation-driven
resonance occurs at $\mu/\mu_{\rm v}\approx 210$. However, evolution
through this resonance is extremely non-adiabatic (see Fig.~\ref{fig-lampn}).
So $\cos\theta_\omega$ is unaffected although $\cos\theta_{\rm IV}$ jumps
from $\approx -1$ to $\approx 1$ immediately after the resonance.
Then $\cos\theta_{\rm IV}$ vanishes at $\mu/\mu_{\rm v}\approx 99$
corresponding to $\omega=3\dot\phi_p-2\mu S_z$ and 
$|\mathbf{H}_{\rm II}|=\omega_n$, but this is not a
resonance (see Sec.~\ref{sec-resnut}). A precession-driven resonance
occurs at $\mu/\mu_{\rm v}\approx 86$ and a second nutation-driven
resonance occurs at $\mu/\mu_{\rm v}\approx 43$. Evolution through
both these resonances is adiabatic (see Fig.~\ref{fig-lampn}). 
Consequently, at $\mu/\mu_{\rm v}<99$,
$\cos\theta_\omega$ oscillate around $\cos\theta_{\rm IV}$, eventually
settling to $\approx -1$ again. There is no net flavor transformation for
$\omega$ in range II. For $\omega/\mu_{\rm v}=-4$ in range III,
$\cos\theta_{\rm IV}$ vanishes at $\mu/\mu_{\rm v}\approx 89$, which
is not a resonance. A precession-driven resonance occurs at
at $\mu/\mu_{\rm v}\approx 69$ and a nutation-driven resonance
occurs at $\mu/\mu_{\rm v}\approx 19$. Evolution is adiabatic 
throughout and there is no net flavor transformation. For 
$\omega/\mu_{\rm v}=-2$ also in range III, the evolution of
$\cos\theta_{\rm IV}$ is similar to that for $\omega/\mu_{\rm v}=-4$.
However, for $\omega/\mu_{\rm v}=-2$,
evolution through the precession-driven resonance at
$\mu/\mu_{\rm v}\approx 51$ is adiabatic while that through the
nutation-driven resonance at $\mu/\mu_{\rm v}\approx 4$
is non-adiabatic (see Fig.~\ref{fig-lampn}). This kind of evolution
applies to $-3<\omega/\mu_{\rm v}<-1$ and results in
significant flavor transformation. 

In summary, evolution of $\mathbf{s}_\omega$ can be understood
in terms of the resonances it experiences. A resonance does not
affect the net flavor transformation if evolution through it is very
non-adiabatic. Net full flavor conversion
results from adiabatic evolution through an odd number of 
resonances while little net flavor transformation results from
adiabatic evolution through an even number (including zero) of 
resonances. As discussed in Secs.~\ref{sec-respre} and \ref{sec-resnut},
evolution through a resonance driven by either precession or
nutation at $\mu>\mu_{\rm cr}$ is extremely non-adiabatic.
In contrast, evolution through a precession-driven resonance at 
$\mu<\mu_{\rm cr}$ is essentially always adiabatic. 
We introduce a parameter $\mu_{\rm tr}$ to discuss the adiabaticity of
evolution through a nutation-driven resonance at $\mu<\mu_{\rm cr}$:
the evolution is adiabatic (non-adiabatic) when such a resonance 
occurs at $\mu_{\rm tr}<\mu<\mu_{\rm cr}$ ($\mu<\mu_{\rm tr}$). 
We choose $\mu_{\rm tr}$ to
correspond to an adiabaticity parameter $\lambda_n=0.5$.
For the neutrino gyroscope in Fig.~\ref{fig-theta}, 
$\mu_{\rm tr}\approx 12\mu_{\rm v}$ and $\mathbf{s}_\omega$ with
$\omega\approx -3\mu_{\rm v}$ goes through a
nutation-driven resonance at $\mu=\mu_{\rm tr}$
(see Figs.~\ref{fig-wres} and \ref{fig-lampn}). 

\begin{figure*}
\includegraphics*[width=6cm, angle=-90]{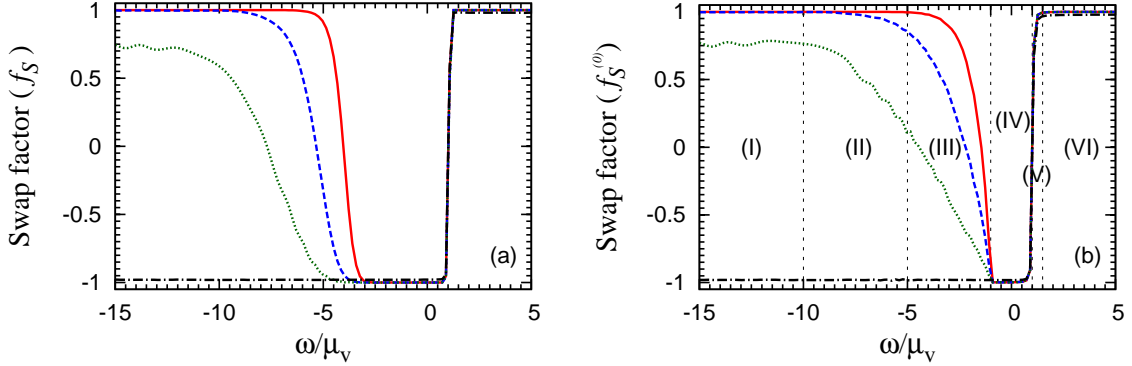}
\caption{Comparison of the swap factors
$f_S(\omega,t)$ (a) and $f_S^{(0)}(\omega,t)$ (b)
at $r=250$~km for the system initially consisting of
$\nu_e$ and $\bar\nu_e$ only as in Fig.~\ref{fig-swapcomp}.
The solid, dashed, dotted, and dot-dashed curves are
for $\tilde\theta_{\rm v}=10^{-5}$, $10^{-3}$, $10^{-2}$, and
$10^{-1}$, respectively. Ranges I--VI of $\omega$ shown
in (b) are based on occurrences of precession-driven and
nutation-driven resonances (calculated for 
$\tilde\theta_{\rm v}=10^{-5}$ but valid for 
$\tilde\theta_{\rm v}\ll1$).\label{fig-swap2}}
\end{figure*}

Now the swap factor shown as the dashed curve in 
Fig.~\ref{fig-swapcomp} can be understood
based on the above discussion and the occurrences of resonances
listed in the beginning of this subsection and
shown in Figs.~\ref{fig-evolres} and \ref{fig-evolnp}. This curve is re-plotted
as the solid curve in Fig.~\ref{fig-swap2}b with ranges I to VI for $\omega$
indicated. Recall that $f_S^{(0)}\approx-1$ corresponds to nearly
full flavor transformation. This applies to $\omega$ in range IV, for which
there is only a single precession-driven resonance at
$\mu<\mu_{\rm cr}$ and evolution through this resonance is adiabatic.
A variety of evolution results in $f_S^{(0)}\approx1$ corresponding to
little net flavor transformation for $\omega$ in ranges I, II, V, and VI:
non-adiabatic evolution through a single precession-driven resonance
at $\mu>\mu_{\rm cr}$ (I), non-adiabatic evolution through a nutation-driven 
resonance at $\mu>\mu_{\rm cr}$ followed by adiabatic evolution through a 
precession-driven resonance and a second nutation-driven resonance at 
$\mu<\mu_{\rm cr}$ (II), adiabatic evolution through two precession-driven
resonances at $\mu<\mu_{\rm cr}$ (V), and adiabatic evolution with
no resonance (VI). A precession-driven resonance
and a nutation-driven resonance occur at $\mu<\mu_{\rm cr}$ for
$\omega$ in range III and evolution through the precession-driven resonance
is always adiabatic. However, evolution through the nutation-driven resonance
is adiabatic only for $\omega<-3\mu_{\rm v}$ in this range and becomes
more and more non-adiabatic as $\omega$ increases above 
$\sim-3\mu_{\rm v}$. Consequently, a transition from $f_S^{(0)}\approx 1$ 
towards $-1$ occurs at $\omega\sim -3\mu_{\rm v}$ in range III.

\subsection{Application to System of Neutrinos with Continuous Spectra
\label{sec-fullspec}}
As discussed in Sec.~\ref{sec-gyromf}, the total angular momentum of 
the neutrino gyroscope, which we denote as $\mathbf{S}^{(0)}$ again
for clarity, approximates the mean field $\mathbf{S}$ of the NFIS's in 
the system of neutrinos with continuous spectra as specified in 
Sec.~\ref{sec-nueanue}. The oscillations of $S_\perp^{(0)}$ and
$S_\perp$ shown in Fig.~\ref{fig-sperp} reflect the nutation of the
gyroscope. It can be seen from this figure that large deviations of
$\mathbf{S}^{(0)}$ from $\mathbf{S}$ occur only at
$\mu/\mu_{\rm v}<20$, where the nutation amplitude of $\mathbf{S}$ 
rapidly decreases. In addition, $S_\perp$ sharply drops at
$\mu/\mu_{\rm v}<2$. The small nutation amplitude of
$\mathbf{S}$ at $\mu/\mu_{\rm v}<20$ affects the adiabaticity of
evolution through the nutation-driven resonance for $\omega$
in range III (see top left panels of Figs.~\ref{fig-evolres} and \ref{fig-evolnp}). 
In fact, the evolution is
extremely non-adiabatic for $-3<\omega/\mu_{\rm v}<-1$. 
On the other hand, evolution through the precession-driven resonance 
is adiabatic for these values of $\omega$, which results in
net full flavor conversion.
Thus, compared with the results based on 
$\mathbf{S}^{(0)}$, the swap factor is $\approx -1$ for
a wider range of $\omega$ as shown by the solid curve
in Fig.~\ref{fig-swap2}a. 
In principle, the sharp decrease of $S_\perp$ at 
$\mu/\mu_{\rm v}<2$ could affect the adiabaticity of evolution
through the precession-driven resonance at the lower $\mu$
for $\omega$ in range V (see Fig.~\ref{fig-evolres}). However,
in practice this has little effect (see the solid curve in
Fig.~\ref{fig-swap2}a) as
resonances at such low values of $\mu$ only occur for a
very narrow range of $\omega$ and adiabaticity is
affected for an even narrower range of $\omega$.

To further illustrate how adiabaticity of evolution through
precession-driven and nutation-driven resonances affect
net flavor transformation, we increase $\tilde\theta_{\rm v}$
from $10^{-5}$ to $10^{-3}$, $10^{-2}$, and $10^{-1}$,
respectively. For a larger $\tilde\theta_{\rm v}$, 
$S_\perp$ in the sleeping-top regime of
$\mu>\mu_{\rm cr}$ is larger as the initial $\theta_p$ of
the neutrino gyroscope becomes larger 
[see Eq.~(\ref{eq-sperp}) and Appendix~\ref{app-rnu}]. 
On the other hand, the nutation amplitude $\eta$ is smaller at 
$\mu<\mu_{\rm cr}$ as it grows less at $\mu\sim\mu_{\rm cr}$
due to a shorter nutation period $\sim2\pi/\omega_{n,{\rm cr}}$
for a larger $\tilde\theta_{\rm v}$ [see Eq.~(\ref{eq-wnc})].
Consequently, evolution through a precession-driven resonance at 
$\mu>\mu_{\rm cr}$ becomes less non-adiabatic while that through
a nutation-driven resonance at $\mu<\mu_{\rm cr}$ 
becomes more non-adiabatic. The former effect becomes
quite large for $\tilde\theta_{\rm v}=10^{-2}$ as partial flavor
conversion occurs for $\omega$ in range I (dotted curves in
Fig.~\ref{fig-swap2}), while the latter effect is already significant
for $\tilde\theta_{\rm v}=10^{-3}$  (dashed curves in
Fig.~\ref{fig-swap2}) as more flavor transformation
occurs for $\omega$ in ranges II and III relative to the case of 
$\tilde\theta_{\rm v}=10^{-5}$ (solid curves in Fig.~\ref{fig-swap2}).
For $\tilde\theta_{\rm v}=10^{-1}$, $S_\perp$ is sufficiently
large initially and $\eta$ remains small at all $\mu$.
Consequently, evolution through precession-driven resonances 
is adiabatic for $\omega$ in ranges I to V 
while nutation-driven resonances 
have no effect on the net flavor transformation. This can be seen
from the dot-dashed curves in Fig.~\ref{fig-swap2}, which show
that net full flavor conversion occurs for $\omega$ in ranges I to IV
with a single precession-driven resonance but there is no net
flavor transformation for $\omega$ in ranges V and VI with two
and zero precession-driven resonances, respectively.

\section{Collective Neutrino Oscillations in Supernovae}\label{sec-snnu}
\begin{figure*}
\includegraphics*[width=16cm, angle=-90]{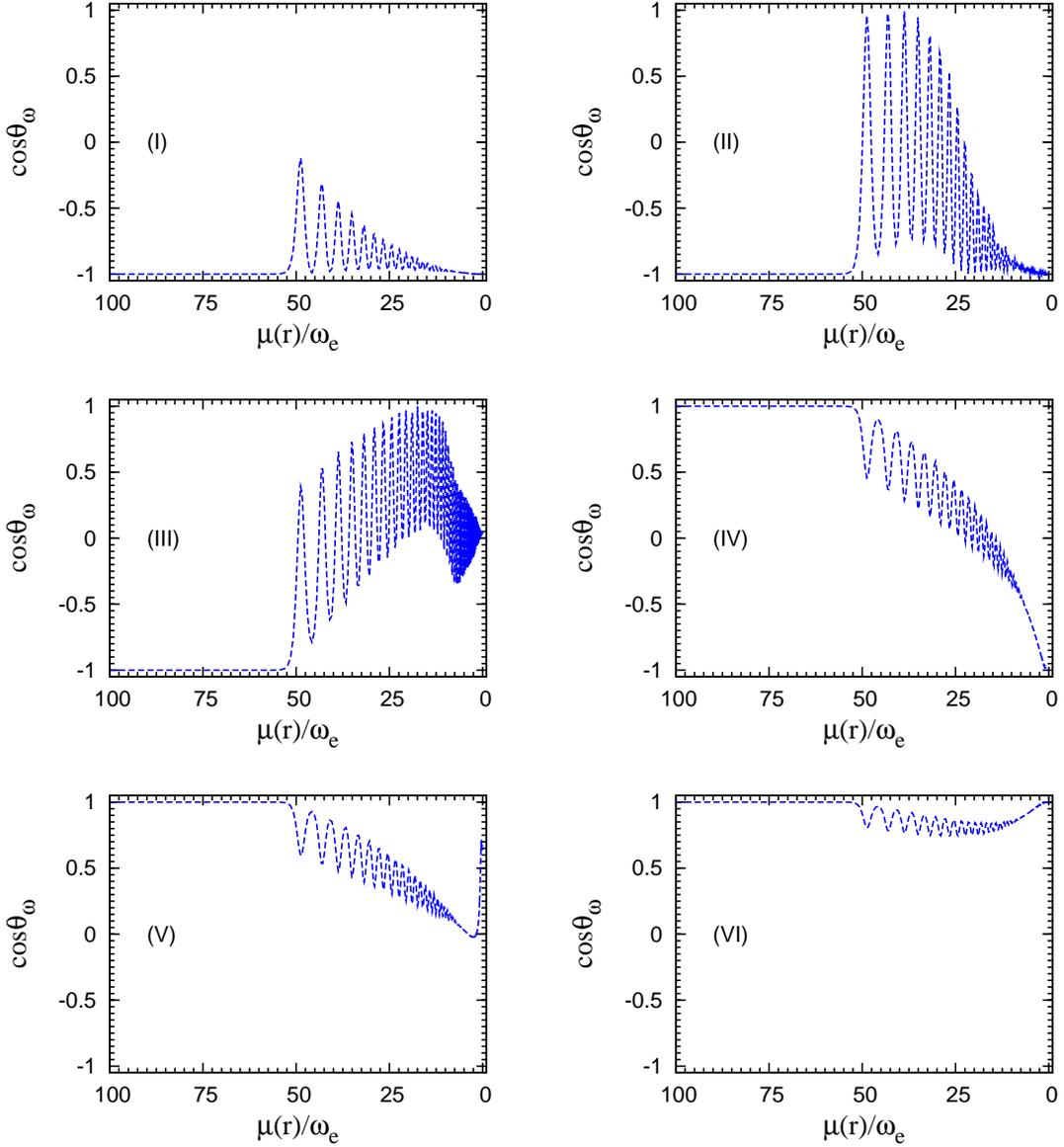}
\caption{Example evolution of $\cos\theta_\omega$ as a
function of $\mu(r)/\omega_e$ for the system shown in Figs.~\ref{fig-splits}
and \ref{fig-swaps}
for $2\omega/\delta m^2=-0.90$ (I), $-0.50$ (II), 
$-0.32$ (III), 0.02 (IV), 0.14 (V), and 0.50 (VI) MeV$^{-1}$, respectively.
Note the similarity to the evolution shown in Figs.~\ref{fig-evolres} and
\ref{fig-evolnp} for the six ranges of $\omega$ based on occurrences of
precession-driven and nutation-driven resonances.\label{fig-evol}}
\end{figure*}

In this section we consider the system of neutrinos exhibiting the spectral
swaps shown in Figs.~\ref{fig-splits} and \ref{fig-swaps}. As described in 
Sec.~\ref{sec-intro}, this system initially consists of $\nu_e$, $\bar\nu_e$,
$\nu_x$, and $\bar\nu_x$ with continuous spectra. In addition, the initial
number densities of $\nu_e$ and $\bar\nu_e$ are significantly larger than
those of $\nu_x$ and $\bar\nu_x$. We first show that the swap factor 
shown in Fig.~\ref{fig-swaps} can be understood in terms of the six
different kinds of flavor evolution that have been
discussed in Secs.~\ref{sec-evolres}
and \ref{sec-fullspec} for the system initially consisting of $\nu_e$ and 
$\bar\nu_e$ only. The evolution of $\cos\theta_\omega$ corresponding
to Fig.~\ref{fig-swaps} is shown in Fig.~\ref{fig-evol} for 
$2\omega/\delta m^2=-0.90$, $-0.50$, $-0.32$, 0.02, 0.14, and 
0.50~MeV$^{-1}$,
respectively. It can be seen that these six kinds of evolution are very
similar to those shown in Figs.~\ref{fig-evolres} and \ref{fig-evolnp} 
for the six ranges of $\omega$ discussed in Secs.~\ref{sec-evolres}.
As the comparison of $\mathbf{S}^{(0)}$ and $\mathbf{S}$ shown in
Fig.~\ref{fig-mfs} for the system of four initial neutrino species is
similar to that shown in Fig.~\ref{fig-sperp} for the system of two
initial neutrino species, the differences between the evolution based
on $\mathbf{S}^{(0)}$ and $\mathbf{S}$ are also similar to those
discussed in Sec.~\ref{sec-fullspec}. 
Therefore, we conclude that the flavor evolution
of a system with initial number densities of $\nu_e$ and $\bar\nu_e$
significantly larger than those of $\nu_x$ and $\bar\nu_x$ can be
understood in terms of the resonances driven by precession and
nutation of a neutrino gyroscope.

Next we consider the flavor evolution of the system exhibiting
the spectral swaps in Figs.~\ref{fig-splits} and \ref{fig-swaps} by relaxing
the single-angle approximation used to produce these results. In the
so-called ``multi-angle'' approximation, neutrinos are emitted from the
neutrino sphere with equal probability in the forward directions, which
are defined to be $0\leq\theta_{\rm em}\leq\pi/2$.
Here $\theta_{\rm em}$ is the angle with respect to the radial direction
at the point of emission. Under the multi-angle approximation, 
an NFIS can be specified by the corresponding neutrino
energy and emission angle
in terms of $\omega$ and $\epsilon\equiv\cos\theta_{\rm em}$. The
evolution of $\mathbf{s}_{\omega,\epsilon}$ is governed by
\begin{equation}\label{eq-nfis_m}
\frac{d}{dr}\mathbf{s}_{\omega,\epsilon}=\mathbf{s}_{\omega,\epsilon}\times
\left\{\frac{\omega}{D_\epsilon(r)}\mathbf{H}_{\rm v}-2\mu(R_\nu)\frac{R_\nu^2}{r^2}
\int_{-\infty}^{\infty}g(\omega')d\omega'\int_{0}^{1}\mathbf{s}_{\omega',\epsilon'}
\left[\frac{1}{D_\epsilon(r) D_{\epsilon^\prime}(r)}-1\right]\epsilon'd\epsilon'\right\},
\end{equation}
where $\mu(R_\nu)=2\sqrt{2}G_Fn_{\nu_e}(R_\nu)$,
$n_{\nu_e}(R_\nu)$ is given by Eq.~(\ref{eq-nr}) for $r=R_\nu$, and
\begin{equation}
D_\epsilon(r)\equiv\sqrt{1-(1-\epsilon^2)R_\nu^2/r^2}.
\end{equation}
Equation~(\ref{eq-nfis_m}) reduces to Eq.~(\ref{eq-nfis3}) under the single-angle
approximation that $\mathbf{s}_{\omega,\epsilon}(r)=
\mathbf{s}_{\omega,\epsilon=1}(r)$.

\begin{figure*}
\includegraphics*[width=5.5cm, angle=-90]{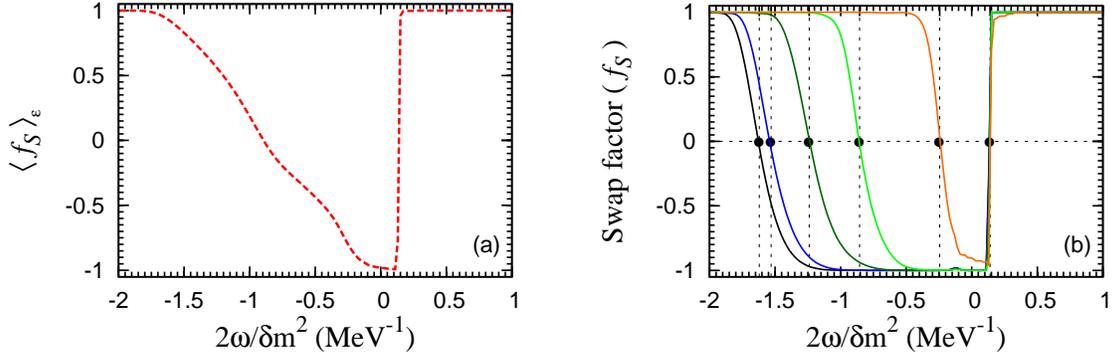}
\caption{Angle-averaged swap factor 
$\langle f_S(\omega,\epsilon,r_f)\rangle_\epsilon$ (a)
and angle-specific swap factors
$f_S(\omega,\epsilon,r_f)$ (b) at $r_f=250$~km
for the system with the same neutrino emission
parameters as for Fig.~\ref{fig-swaps} but in
multi-angle approximation. From left to right,
the curves for $\omega<0$ in (b) are for
$\epsilon=10^{-3}$, 0.24, 0.49, 0.74, and 0.99, respectively.
The sharp changes in $f_S(\omega,\epsilon,r_f)$
occur at $2\omega/\delta m^2=-1.63$, $-1.54$, $-1.24$, $-0.86$,
$-0.24$, and 0.14~MeV$^{-1}$, respectively, for which 
$f_S(\omega,\epsilon,r_f)=0$ (filled circles).
\label{fig-swapmulti}}
\end{figure*}

Using the same neutrino emission parameters as for Figs.~\ref{fig-splits} and
\ref{fig-swaps}, we follow the flavor evolution of the system with initial number
densities of $\nu_e$ and $\bar\nu_e$ significantly larger than those of
$\nu_x$ and $\bar\nu_x$ under the multi-angle approximation.
The angle-averaged swap factor 
$\langle f_S(\omega,\epsilon,r)\rangle_\epsilon$ at $r=r_f=250$~km
is shown as a function of $2\omega/\delta m^2$ in Fig.~\ref{fig-swapmulti}a.
This average factor is defined as
\begin{equation}
\langle f_S(\omega,\epsilon,r)\rangle_\epsilon\equiv
\frac{\int_0^1f_S(\omega,\epsilon,r)[\epsilon/D_\epsilon(r)]d\epsilon}
{\int_0^1[\epsilon/D_\epsilon(r)]d\epsilon}=
\frac{\int_0^1f_S(\omega,\epsilon,r)[\epsilon/D_\epsilon(r)]d\epsilon}
{\left[1-\sqrt{1-R_\nu^2/r^2}\right]r^2/R_\nu^2},
\end{equation}
and can be used to calculate the effect of flavor transformation on 
e.g., neutrino reaction rates at radius $r$. Compared with the swap factor
shown in Fig.~\ref{fig-swaps} for the single-angle approximation, 
$\langle f_S(\omega,\epsilon,r_f)\rangle_\epsilon$ is identical for $\omega>0$
but shows large deviations for $\omega<0$. On the other hand, 
swap factors $f_S(\omega,\epsilon,r_f)$ for specific values of $\epsilon$ 
shown in Fig.~\ref{fig-swapmulti}b have the same general
structure as shown in Fig.~\ref{fig-swaps}. This can be understood 
from the precession-driven resonance as discussed below.

\begin{figure}
\includegraphics*[width=5.5cm, angle=-90]{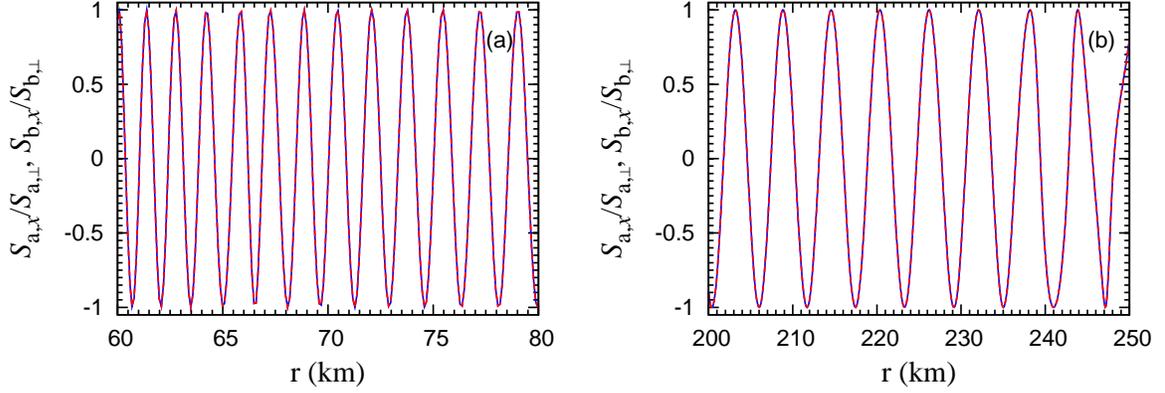}
\caption{Synchronized oscillations of $S_{a,x}/S_{a,\perp}$ and 
$S_{b,x}/S_{b,\perp}$ due to precession for the system shown in 
Fig.~\ref{fig-swapmulti} for $r=60$--80 (a) and 200--250~km (b). 
In either panel two curves are shown
but they are indistinguishable.\label{fig-sabx}}
\end{figure}

\begin{figure}
\includegraphics*[width=10cm, angle=-90]{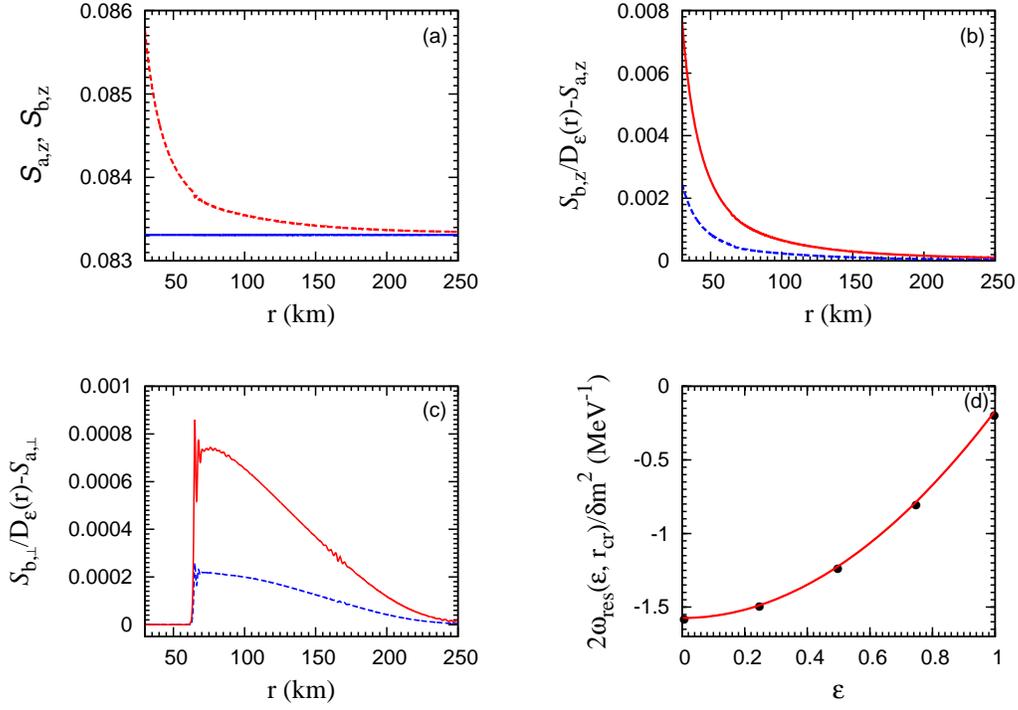}
\caption{Components of the mean field as functions of $r$:
(a) $S_{a,z}$ (solid curve) and $S_{b,z}$ (dashed curve), 
(b) $S_{b,z}/D_\epsilon(r)-S_{a,z}$ for $\epsilon=0$ (solid curve)
and 1 (dashed curve), and (c) $S_{b,\perp}/D_\epsilon(r)-S_{a,\perp}$ 
for $\epsilon=0$ (solid curve) and 1 (dashed curve). (d) The value of
$2\omega_{\rm res}(\epsilon,r_{\rm cr})/\delta m^2$ at 
$r_{\rm cr}\approx 64$~km shown as a function of $\epsilon$. 
From left to right, the filled circles correspond to 
$2\omega_{\rm res}(\epsilon,r_{\rm cr})/\delta m^2\approx-1.57$, 
$-1.50$, $-1.24$, $-0.81$, and $-0.21$~MeV$^{-1}$, respectively.
These values are very close to those values of 
$2\omega/\delta m^2$ that correspond to the
filled circles at $\omega<0$ in Fig.~\ref{fig-swapmulti}b. 
\label{fig-sab}}
\end{figure}

It is convenient to define 
\begin{subequations}
\begin{align}
\mathbf{S}_a\equiv & \int_{-\infty}^{\infty}g(\omega)d\omega
\int_{0}^{1}\mathbf{s}_{\omega,\epsilon}\epsilon d\epsilon,\\
\mathbf{S}_b\equiv & \int_{-\infty}^{\infty}g(\omega)d\omega
\int_{0}^{1}\mathbf{s}_{\omega,\epsilon}\left[\frac{\epsilon}{D_\epsilon(r)}\right]d\epsilon,
\end{align}
\end{subequations} 
and rewrite Eq.~(\ref{eq-nfis_m}) as
\begin{equation}\label{eq-NFIS_m}
\frac{d}{dr}\mathbf{s}_{\omega,\epsilon}=\mathbf{s}_{\omega,\epsilon}\times
\left\{\frac{\omega}{D_\epsilon(r)}\mathbf{H}_{\rm v}-2\mu(R_\nu)\frac{R_\nu^2}{r^2}
\left[\frac{\mathbf{S}_b}{D_\epsilon(r)}-\mathbf{S}_a\right]\right\}.
\end{equation}
Our numerical results show that both $\mathbf{S}_a$ and $\mathbf{S}_b$ precess
around $\mathbf{H}_{\rm v}$ with the same frequency $\dot\phi_p(r)$ at a specific
$r$ for $R_\nu\leq r\lesssim r_f$. This is consistent with the conclusions of
\cite{08082046} based on symmetry arguments.
The synchronized oscillations of
$S_{a,x}/S_{a,\perp}$ and $S_{b,x}/S_{b,\perp}$ due to precession are shown 
for $r=60$--80 and 200--250~km in Figs.~\ref{fig-sabx}a and b, respectively.
Here the $x$-axis is in the plane perpendicular to $\mathbf{H}_{\rm v}$ 
and the subscript ``$\perp$'' denotes the net perpendicular component. 
It can be shown from Eq.~(\ref{eq-NFIS_m}) that 
$S_{a,z}\equiv\mathbf{S}_a\cdot\mathbf{\hat e}_z^{\rm I}$ 
($\mathbf{\hat e}_z^{\rm I}=-\mathbf{H}_{\rm v}$)
is conserved (see Fig.~\ref{fig-sab}a). Components of the ``mean field'' 
$\mathbf{S}_b/D_\epsilon(r)-\mathbf{S}_a$ are shown as functions of $r$
for $\epsilon=0$ (solid curve) and 1 (dashed curve) in Figs.~\ref{fig-sab}b and c.
Clearly, the mean field experienced by $\mathbf{s}_{\omega,\epsilon}$ is different 
for different $\epsilon$. It can also be seen from Fig.~\ref{fig-sab}c that the
perpendicular component $S_{b,\perp}/D_\epsilon(r)-S_{a,\perp}$ increases
sharply at $r_{\rm cr}\approx 64$~km just as $S_\perp$ does at the critical point
in the single-angle approximation (see Fig.~\ref{fig-mfs}). However, in contrast
to the single-angle approximation, nutation of the mean field
is damped out very quickly in the multi-angle approximation, and 
therefore, can be neglected.

Based on the above discussion, we only need to consider the precession-driven
resonance in explaining the swap factor $f_S(\omega,\epsilon,r_f)$.
A precession-driven resonance occurs when the $z$-component of the net effective 
field interacting with $\mathbf{s}_{\omega,\epsilon}$ vanishes in the
co-precessing frame (see Sec.~\ref{sec-respre}). 
It can be seen from Eq.~(\ref{eq-NFIS_m}) that
$\mathbf{s}_{\omega,\epsilon}$ goes through a precession-driven resonance when
\begin{equation}
\omega=\omega_{\rm res}(\epsilon,r)\equiv
D_\epsilon(r)\left\{\dot\phi_p(r)-2\mu(R_\nu)\frac{R_\nu^2}{r^2}
\left[\frac{S_{b,z}}{D_\epsilon(r)}-S_{a,z}\right]\right\}.
\end{equation}
However, evolution through a resonance at $r<r_{\rm cr}$ is extremely
non-adiabatic and does not result in any net flavor transformation.
Thus, we expect that net full conversion [$f_S(\omega,\epsilon,r_f)=-1$]
occurs only for
$\omega_{\rm res}(\epsilon,r_{\rm cr})<\omega<\omega_{\rm res}(\epsilon,r_f)$.
At $r=r_{\rm cr}$, $\omega_{\rm res}(\epsilon,r_{\rm cr})<0$ for $0\leq\epsilon\leq 1$
and is larger for a larger $\epsilon$ (more radial trajectory, see Fig.~\ref{fig-sab}d).
At $r=r_f$, $D_\epsilon(r)\approx 1$ and 
$S_{b,z}/D_\epsilon(r)-S_{a,z}$ becomes very small (see Fig.~\ref{fig-sab}b), 
so $\omega_{\rm res}(\epsilon,r_f)\approx\dot\phi_p(r_f)$ and is nearly
independent of $\epsilon$. Using our numerical results for $\dot\phi_p(r)$,
$S_{a,z}$, and $S_{b,z}$, we obtain
${2\omega_{\rm res}(\epsilon,r_{\rm cr})}/{\delta m^2}\approx -1.57$, $-1.50$,
$-1.24$, $-0.81$, and $-0.21$~MeV$^{-1}$ for $\epsilon=10^{-3}$, 0.24, 0.49,
0.74, and 0.99, respectively, and 
${2\omega_{\rm res}(\epsilon,r_f)}/{\delta m^2}\approx 0.14$ MeV$^{-1}$ for
$0\leq\epsilon\leq 1$. These results are in excellent agreement with
Figure~\ref{fig-swapmulti}b.

\section{Conclusions}
\label{sec-conc}
Using a system initially consisting of $\nu_e$ and $\bar\nu_e$ with the
same energy spectrum but different number densities, we have shown that
flavor evolution of this system in the single-angle approximation
can be understood in terms of the response of individual NFIS's to the mean field, 
which is very well approximated by the total angular momentum of a
neutrino gyroscope. The evolution of an NFIS is
governed by two types of resonances driven by precession and nutation of
the gyroscope, respectively. A resonance does not affect the net flavor
transformation if evolution through it is extremely non-adiabatic.
Nearly full flavor conversion occurs following
adiabatic evolution through an odd number of
resonances but there is no net flavor transformation following adiabatic
evolution through an even number (including zero) of resonances.
The detailed results on NFIS evolution are presented in Figs.~\ref{fig-evolres}, 
\ref{fig-evolnp}, and \ref{fig-swap2}, 
and discussed in Secs.~\ref{sec-evolres} and \ref{sec-fullspec}.

We have also shown that the above results for the system of two initial neutrino
species can be extended to a system of four species with the initial
number densities of $\nu_e$ and $\bar\nu_e$ significantly larger than those of
$\nu_x$ and $\bar\nu_x$. Further, we find that when the multi-angle 
approximation is adopted instead of the single-angle approximation,
nutation of the mean field is quickly damped out and can be neglected.
In contrast, precession-driven resonances still govern the evolution of
NFIS's with different energy and emission angles just as in the single-angle
approximation. These results are presented and discussed in Sec.~\ref{sec-snnu}.

In conclusion, we have presented a detailed analysis of collective 
neutrino oscillations
in supernovae for the case where the initial number densities of 
$\nu_e$ and $\bar\nu_e$ are significantly larger than those of
$\nu_x$ and $\bar\nu_x$. We note that some earlier works
(e.g., \cite{07094641}) and two recent
studies \cite{1103.2891,1103.5302} 
have similar goals to ours but used very different methods.
Our approach is mostly pedagogical and analytic. It is our hope that
along with other parallel efforts, we have provided some insights into the 
seemingly complicated yet fascinating 
phenomena of collective neutrino oscillations.

\begin{acknowledgments}
Y.-Z.Q. thanks Joe Carlson, John Cherry, Huaiyu Duan, and George Fuller
for fruitful collaboration on collective neutrino oscillations in supernovae.
This work was supported in part by the US DOE under DE-FG02-87ER40328 at UMN.
\end{acknowledgments}

\appendix
\section{Parameters of the Neutrino Gyroscope at a specific $\mu$\label{app-gyro}}
In addition to the procedure given in Sec.~\ref{sec-vnnu},
the parameters of the neutrino gyroscope at a specific $\mu$ can be
obtained using the ``pure-precession'' ansatz (see also discussion in
\cite{0703776}), which assumes 
that $\mathbf{s}_1$ and $\mathbf{s}_2$ associated with the gyroscope
always stay in the same plane as 
$\mathbf{H}_{\rm v}=-\mathbf{\hat e}_z^{\rm I}$
and precess with the same angular 
velocity $\dot\phi_p\mathbf{\hat e}_z^{\rm I}$. More specifically,
we can use the Euler angles in Frame I to write
\begin{subequations}
\begin{align}
\mathbf{s}_1&=\frac{1}{2}(\sin\theta_1\sin\phi_1\mathbf{\hat e}_x^{\rm I}
-\sin\theta_1\cos\phi_1\mathbf{\hat e}_y^{\rm I}+
\cos\theta_1\mathbf{\hat e}_z^{\rm I}),\\
\mathbf{s}_2&=\frac{1}{2}(\sin\theta_2\sin\phi_2\mathbf{\hat e}_x^{\rm I}
-\sin\theta_2\cos\phi_2\mathbf{\hat e}_y^{\rm I}+
\cos\theta_2\mathbf{\hat e}_z^{\rm I}).
\end{align}
\end{subequations}
The ansatz assumes that $\dot\theta_1=\dot\theta_2=0$,
$\dot\phi_1=\dot\phi_2=\dot\phi_p$, and
$\phi_2-\phi_1=\pi$ (the last relation can be seen 
from the initial configuration at the neutrino
sphere with $\mathbf{s}_1=-\mathbf{s}_2=\mathbf{\hat e}_z^{\rm f}/2$
corresponding to $\theta_1=2\tilde\theta_{\rm v}$ and
$\theta_2=\pi-2\tilde\theta_{\rm v}$).
Using this ansatz along with conservation of $S_z$ and 
Eqs.~(\ref{eq-s1}) and (\ref{eq-s2}), we obtain
\begin{subequations}
\begin{align}
\cos\theta_1+\alpha\cos\theta_2&=
(1-\alpha)\cos2\tilde\theta_{\rm v},\\
\dot\phi_p\sin\theta_1&=\mu_{\rm v}\sin\theta_1+
\frac{\alpha\mu}{2}\sin(\theta_1+\theta_2),\\
\dot\phi_p\sin\theta_2&=-\mu_{\rm v}\sin\theta_2+
\frac{\mu}{2}\sin(\theta_1+\theta_2).
\end{align}
\end{subequations}
For any specific $\mu$,
the above equations can be solved to give
$\theta_1$, $\theta_2$, and $\dot\phi_p$. Then we can
calculate the corresponding $\sigma$, $Q$, and $\theta_p$
using $\theta_1$ and $\theta_2$.
The above procedure gives $\sigma$, $Q$, $\theta_p$,
and $\dot\phi_p$ for the gyroscope at a specific $\mu$
that are indistinguishable from those obtained by
the procedure discussed in Sec.~\ref{sec-vnnu}.

\section{Initial Conditions for the Neutrino Gyroscope\label{app-rnu}}
The initial conditions for the neutrino gyroscope at the neutrino sphere are
\begin{subequations}
\begin{align}
\mathbf{S}(0)&=\frac{1-\alpha}{2}\mathbf{\hat e}_z^{\rm f},\\
\mathbf{Q}(0)&=\frac{1+\alpha}{2}\mathbf{\hat e}_z^{\rm f}+
\frac{\mu_{\rm v}}{\mu}\mathbf{H}_{\rm v}.
\end{align}
\end{subequations}
Noting that $\mathbf{H}_{\rm v}=-\sin2\tilde\theta_{\rm v}
\mathbf{\hat e}_x^{\rm f}-\cos2\tilde\theta_{\rm v}\mathbf{\hat e}_z^{\rm f}
=-\mathbf{\hat e}_z^{\rm I}$, we obtain from the above equations
\begin{subequations}
\begin{align}
S_z&=\frac{1-\alpha}{2}\cos2\tilde\theta_{\rm v},\\
Q&=\left[\left(\frac{1+\alpha}{2}-\frac{\mu_{\rm v}}{\mu}
\cos2\tilde\theta_{\rm v}\right)^2+\left(\frac{\mu_{\rm v}}{\mu}
\sin2\tilde\theta_{\rm v}\right)^2\right]^{1/2},\\
\sigma&=\frac{1-\alpha}{2Q}\left(\frac{1+\alpha}{2}-
\frac{\mu_{\rm v}}{\mu}\cos2\tilde\theta_{\rm v}\right).
\end{align}
\end{subequations}

In terms of the dynamic variables $\theta$ and $\phi$, the
initial conditions for the neutrino gyroscope can be chosen as
$\dot\theta_0=0$, $\phi_0=0$, and
\begin{subequations}
\begin{align}
\cos\theta_0&=\frac{1}{Q}\left(\frac{1+\alpha}{2}
\cos2\tilde\theta_{\rm v}-\frac{\mu_{\rm v}}{\mu}\right),
\label{eq-theta0}\\
\dot\phi_0&=\mu_{\rm v}\frac{1-\alpha}{1+\alpha},
\label{eq-dotphi0}
\end{align}
\end{subequations}
where the subscript ``0'' indicates the initial moment $t=0$
and Eq.~(\ref{eq-dotphi0}) is obtained from Eq.~(\ref{eq-qg})
at $t=0$. For a constant $\mu$,
the precession frequency at $\theta=\theta_p$, where
$V_{\rm eff}(\theta)$ reaches its minimum,
is given by Eq.~(\ref{eq-pre}) as
\begin{equation}
\dot\phi_p=\frac{\mu\sigma\pm
\sqrt{(\mu\sigma)^2-4\mu\mu_{\rm v}Q\cos\theta_p}}{2\cos\theta_p}
\approx\begin{cases}
(\mu\sigma/\cos\theta_p)-(\mu_{\rm v}Q/\sigma),\\
\mu_{\rm v}Q/\sigma,
\end{cases}
\end{equation}
where the approximate equalities apply for $\mu_{\rm v}/\mu\ll 1$ 
with the upper and lower expressions corresponding to the plus and
minus signs in front of the square root, respectively.
It can be shown that the upper expression of $\dot\phi_p$ is unphysical
as it cannot satisfy conservation of $S_z$. For the
physical value of $\dot\phi_p\approx\mu_{\rm v}Q/\sigma$,
conservation of $S_z$ gives
\begin{equation}
\theta_p\approx 2\tilde\theta_{\rm v}\left[1+
\frac{2(1+\alpha)}{(1-\alpha)^2}\frac{\mu_{\rm v}}{\mu}\right].
\end{equation}
The above expression of $\theta_p$ is to the first order in
$\mu_{\rm v}/\mu$ and $\tilde\theta_{\rm v}$. 
To the same order, we have
\begin{equation}
\theta_0\approx 2\tilde\theta_{\rm v}\left(
1+\frac{2}{1+\alpha}\frac{\mu_{\rm v}}{\mu}\right).
\end{equation}
The amplitude of nutation around $\theta=\theta_p$ is then
\begin{equation}\label{eq-etaapp}
\eta=\theta_p-\theta_0\approx
\frac{16\alpha\tilde\theta_{\rm v}}{(1+\alpha)(1-\alpha)^2}
\frac{\mu_{\rm v}}{\mu}.
\end{equation}

The number density of $\nu_e$ at the neutrino sphere is
\begin{equation}
n_{\nu_e}(R_\nu)=\frac{L_{\nu_e}}{4\pi R_\nu^2\langle E_{\nu_e}\rangle}
=1.66\times 10^{32}\left(\frac{L_{\nu_e}}{10^{51}\ {\rm erg/s}}\right)
\left(\frac{10\ {\rm km}}{R_\nu}\right)^2
\left(\frac{10\ {\rm MeV}}{\langle E_{\nu_e}\rangle}\right)\ {\rm cm}^{-3}.
\end{equation}
For $\mu=2\sqrt{2}G_Fn_{\nu_e}(R_\nu)$,
\begin{equation}
\frac{\mu_{\rm v}}{\mu}=5.92\times 10^{-6}
\left(\frac{\delta m^2}{3\times 10^{-3}\ {\rm eV}^2}\right)
\left(\frac{10\, {\rm MeV}}{E}\right)
\left[\frac{10^{32}\ {\rm cm}^{-3}}{n_{\nu_e}(R_\nu)}\right].
\end{equation}
Taking $\tilde\theta_{\rm v}=10^{-5}$, $\delta m^2=3\times 10^{-3}$~eV$^2$,
$2\mu_{\rm v}/\delta m^2=1/9$~MeV$^{-1}$, $\alpha=2/3$, and
$n_{\nu_e}(R_\nu)=1.66\times 10^{32}$~cm$^{-3}$,
we have $\eta\approx 2.28\times 10^{-9}$.

\section{The Neutrino Gyroscope at the Critical Point\label{app-critical}}
The critical point at $\mu=\mu_{\rm cr}$ separates 
the evolution of the neutrino gyroscope into two regimes: the sleeping-top 
regime with essentially pure precession but little nutation at 
$\mu\gtrsim\mu_{\rm cr}$ 
and the other with both precession and nutation at $\mu<\mu_{\rm cr}$.
As $\theta\ll 1$ in the sleeping-top regime, we derive $\mu_{\rm cr}$
assuming $\tilde\theta_{\rm v}=0$. Let the gyroscope start with
$\theta=0$ at a constant $\mu$. Its fixed parameters are
\begin{subequations}
\begin{align}
S_z&=\sigma=\frac{1-\alpha}{2},\\
Q&=\frac{1+\alpha}{2}-\frac{\mu_{\rm v}}{\mu},\\
E_{\rm gyro}&=\frac{\mu}{2}\sigma^2+\mu_{\rm v}Q.
\end{align}
\end{subequations}
The motion of the gyroscope is governed by
\begin{subequations}
\begin{align}
\frac{\dot\phi}{\mu}\sin^2\theta&=\sigma(1-\cos\theta),\label{eq-szst}\\
\frac{1}{2\mu}(\dot\theta^2+\dot\phi^2\sin^2\theta)&=
\mu_{\rm v}Q(1-\cos\theta),\label{eq-est}
\end{align}
\end{subequations}
which are obtained by rewriting Eqs.~(\ref{eq-sz}) and (\ref{eq-etp}).
Assuming that $\theta>0$ is allowed, we can find $\theta_{\rm max}$,
the maximum value of $\theta$, by setting $\dot\theta=0$ in
Eq.~(\ref{eq-est}). Combining the resulting equation with
Eq.~(\ref{eq-szst}), we obtain
\begin{equation}
\cos\theta_{\rm max}=\frac{\mu\sigma^2}{2\mu_{\rm v}Q}-1.
\end{equation}
It can be seen that when $\mu\sigma^2\geq 4\mu_{\rm v}Q$,
the above equation has no solution for $\theta_{\rm max}>0$.
Thus, the gyroscope remains in its initial vertical position 
($\theta=0$) for $\mu\sigma^2\geq 4\mu_{\rm v}Q$. For a
gyroscope starting at $\mu\gg\mu_{\rm v}$, this condition
corresponds to $\mu\geq\mu_{\rm cr}$, where
\begin{equation}
\mu_{\rm cr}\equiv\frac{4\mu_{\rm v}}{(1-\sqrt{\alpha})^2}.
\end{equation}

Expanding
$S_z=(1-\alpha)\cos2\tilde\theta_{\rm v}/2$ to the leading order in
$\tilde\theta_{\rm v}$, we obtain the pure-precession solution at
$\mu=\mu_{\rm cr}$ from
Eqs.~(\ref{eq-pre}), (\ref{eq-sigq}), (\ref{eq-sz2}), and (\ref{eq-qst2})
(setting $\eta=0$ in the last equation):
\begin{subequations}\label{eq-tpexpand}
\begin{align}
\sigma_{\rm cr}&\approx\sigma^{(0)}\left[1-\frac{(1-\sqrt{\alpha})^2}
{8\sqrt{2\alpha}}\theta_{p,{\rm cr}}^3\right],\\
Q_{\rm cr}&\approx Q^{(0)}\left[1+\frac{(1-\sqrt{\alpha})^2}
{4\sqrt{2\alpha}}\theta_{p,{\rm cr}}^3\right],\\
\theta_{p,{\rm cr}}&\approx\frac{2\sqrt{2}\alpha^{1/6}}{(1+\sqrt{\alpha})^{2/3}}
\tilde\theta_{\rm v}^{2/3},\\
\dot\phi_{p,{\rm cr}}&\approx\dot\phi_p^{(0)}\left(1-\frac{\theta_{p,{\rm cr}}}{\sqrt{2}}
+\frac{1+\alpha}{4\sqrt{2}}\theta_{p,{\rm cr}}^2\right),
\end{align}
\end{subequations}
where $\sigma^{(0)}=(1-\alpha)/2$, 
$Q^{(0)}=(1+\alpha)/2-\mu_{\rm v}/\mu_{\rm cr}$,
and $\dot\phi_p^{(0)}=\mu_{\rm cr}\sigma^{(0)}/2$. 
For $\alpha=2/3$ and $\tilde\theta_{\rm v}=10^{-5}$, we have
$\theta_{p,{\rm cr}}\approx 8.24\times 10^{-4}$,
which agrees with the numerical result very well.
Using the above results at the critical point and Eq.~(\ref{eq-d2vdt}), we obtain
\begin{equation}
\omega_{n,{\rm cr}}\approx\frac{\sqrt{3}}{2}\mu_{\rm cr}\sigma^{(0)}\theta_{p,{\rm cr}}.
\end{equation}
We note that the behavior of the precession and nutation frequencies 
at the critical point as shown in Fig.~\ref{fig-nutpre} is unique to the IH.
In contrast, both the precession and nutation 
frequencies increase smoothly with $\mu$ for the NH.

\bibliography{prd}

\end{document}